\documentclass[]{aa}
\usepackage{natbib} % nota bene: to get the .bbl file for the ArXiv submission, download the project, recompile it on the local machine and add the file to the overleaf project. 
\usepackage{lscape}
\usepackage{txfonts}
\usepackage{graphicx}
\usepackage[switch]{lineno}
\newcommand{\apo}{APOKASC\,3\xspace}
\newcommand{\apok}{APO-K2\xspace}

\newcommand{\num} {$\nu_{\rm max}$\xspace}
\newcommand{\logg} {$\log g$\xspace}
\newcommand{\dnu} {$\Delta\nu$\xspace}

\newcommand{\Gaia} {\textit{Gaia}\xspace}
\newcommand{\gaia} {\textit{Gaia}\xspace}
\newcommand{\GaiaDR} {\textit{Gaia}\,DR3\xspace}

\newcommand{\Kepler} {\textit{Kepler}\xspace}
\newcommand{\kepler} {\textit{Kepler}\xspace}

\def\dnu{$\Delta\nu$\xspace}

\def\dpi{$\Delta\Pi_1$\xspace}
\def\dn1{$\delta\nu_{01}$\xspace}
\def\dn2{$\delta\nu_{02}$\xspace}

\newcommand{\blu}{\textcolor{blue} }

\newcommand{\change}[1]{\textit{\blu{#1}}}

\usepackage[colorlinks=true,
                        linkcolor=cyan,
                        citecolor=cyan, 
                        filecolor=cyan, 
                        urlcolor=cyan]{hyperref}

\begin{document} 

\title{%Seismic inference on stellar and orbital properties of binary systems hosting red giants – 
Tales of stellar and binary coevolution, told by stellar oscillations
}
\subtitle{Binary demographics and their impact on stellar mass, orbits, \\and age estimates in main-sequence and red-giant stars}

\titlerunning{Seismic inference on stellar and orbital properties of binary systems hosting red giants}
\authorrunning{P.\,G.\,Beck}

\author{Paul\,G.\,Beck}%\orcidlink{0000-0003-4745-2242}}

\institute{Instituto de Astrof\'{\i}sica de Canarias, E-38200 La Laguna, Tenerife, Spain; \label{inst:IAC}
\email{paul.beck@iac.es, paul.beck@gmx.at}
\and Departamento de Astrof\'{\i}sica, Universidad de La Laguna, E-38206 La Laguna, Tenerife, Spain \label{inst:ULL}
}

\date{Submitted: April 14, 2025, Accepted: December 8, 2025}

\abstract
%Context.
{Red giants are increasingly used as stellar population tracers due to their well-understood evolution and the availability of asteroseismic observables. However, stellar binarity can alter observable properties and introduce strong biases.}
%Aims.
{We aim to assess a holistic picture of the binary population and evolution in the red-giant phase by characterizing the sample of binaries hosting oscillating red giants from a combination of large asteroseismic, spectroscopic, and astrometric surveys.}
%Methods. 
{We investigated the binary properties of evolved stars in the \apo and \apok catalogs, leveraging asteroseismic constraints and \Gaia DR3 non-single-star solutions. We explored the mass distribution of red-giant binary systems, analyzed the evolution of their binary fraction. We investigated the impact of stellar evolution on the orbital periods ($P_\mathrm{orb}$), eccentricities, radial velocity (RV) amplitudes, and the fractional radius and identifed candidate systems that may have undergone significant interactions.}
%Results. 
{For stars with M\,$\leq$\,1.8M\,$_\odot$, we find binary fractions $\sim$31\% and $\sim$41\% for oscillating and non-oscillating solar-like stars on the main sequence (MS). By using the peak frequency of the oscillation power excess ($\nu_\mathrm{max}$) as luminosity proxy and evolutionary states, we detect a binary attrition of $\sim$69\% and $\sim$81\% on the low- and high-luminosity red-giant branch (RGB) and an additional $\sim$38\% to the red clump (RC), with respect to the MS.  Binaries hosting RC and secondary-clump stars (2RC) are largely depleted at $P_\mathrm{orb}$\,$\lesssim$\,500 and $\lesssim$\,200\,days, respectively. We identify a population of rapidly rotating RC stars in short-period orbits as potential post-common-envelope merger products. {Mass-dependent differences in binary fractions and orbital properties point to stronger binary attrition for stars with M\,$\leq$1.8\,M$_\odot$.}}
%Conclusions.
{Binarity is not the primary cause of reduced oscillation amplitudes in MS solar-like stars. The distinct mass distributions and depletion of short-period binaries in the red-giant phase underscore the impact of stellar expansion and binary interaction on stellar evolution. Helium-core burning RC systems with $P_\mathrm{orb}$\,$\lesssim$\,800 to 1,000 days are likely shaped by past interactions, such as mass transfer or loss, which can lead to significantly biased age estimates if not accounted for. This demonstrates the importance of identifying stellar binarity when using red giants as population tracers.
}

\let\linenumbers\nolinenumbers\nolinenumbers

\maketitle

\linenumbers\modulolinenumbers[5]
 
\section{Introduction}
Understanding how stars evolve in binary systems is crucial for building a holistic picture of the coevolution of stars and the evolution of a binary system as a whole. Binary systems are formed in a wide range of orbital periods and eccentricities, leading to diverse evolutionary pathways depending on the initial masses, separations, and interactions between the components  \citep[][and references therein]{Offner2022, MoeStefano2017}. A star’s internal structure, which is governed by its mass and evolutionary stage \citep[][]{Salaris2005, Kippenhahn2013, Pinsonneault2023}, dictates how it responds to binary interactions \citep[][]{Zahn1977, Zahn2013, Ogilvie2013, Esseldeurs2024}. This interplay not only influences the evolution of the individual stars but also shapes the orbital architecture and the ultimate fate of the system as a whole.

Unless formed through a rare capturing event, binary stars are generally coeval, having formed from the same molecular cloud \citep[][]{Offner2022}, justifying the assumption of an identical initial chemical composition \citep[][]{Torres2010}. 
Therefore, the ages of its stellar components are typically assumed to be similar, within the free-fall time of the star-forming filament of the interstellar cloud from which the binary system has emerged \citep[][]{Offner2022}. At a common distance, differences in luminosity and color can be attributed to differences in evolutionary stage rather than interstellar extinction. 
Given these shared characteristics, the key parameter governing the disparity between the two components is their mass difference, which can be accurately determined from the radial velocities (RV) solution of a double-lined spectroscopic binary (SB2) system. In particular, in the red-giant phase, mass ratios differing from unity by a few percent leads to pronounced differences in the stellar evolution of the two components \citep{Miglio2014, Beck2018Asterix}. Such systems are benchmarks for calibrating stellar models, testing complex physics, and validating model-dependent age estimates \citep{delBurgo2018, Beck2018Asterix, Jorgensen2020, Grossmann2025}.

About half of all solar-like stars (from late F, G, to early K; 0.8$\lesssim$\,M/M$_\odot$\,$\lesssim$\,1.5) on the main sequence (MS) are members of binary or multiple systems \citep{Lada2006, Offner2022, MoeStefano2017}. However, the binary fraction and the distribution of orbital parameters evolve as the stars evolve. The most significant transformations occur during the red-giant branch (RGB) and asymptotic-giant branch (AGB) phases, where the evolving star experiences a dramatic expansion of its radius \citep[e.g.,][]{Serenelli2017TRGB, Hekker2020}. These changes can trigger complex interactions, such as tidal effects, mass transfer, and common-envelope evolution that substantially alter both stellar and orbital properties \citep{Ivanova2013}. Based on stellar population synthesis, \citet{Mazzi2025} estimates that $\sim$1\% of red giants observed by the NASA \Kepler mission \citep{Borucki2010} have undergone mass transfer episodes. Such interactions can rejuvenate a star or make it appear prematurely evolved, potentially biasing age estimates and affecting the inferred age-metallicity relation of the Galaxy  \citep{Mazzi2025}.

Asteroseismology of solar-like oscillators has proven to be a powerful tool to characterize individual stars and stellar ensembles. Solar-like oscillations are stochastically excited by convective motions in the outer layers of stars and are observed from the main sequence (MS) through to the red-giant branch (RGB) \citep[e.g.,][and references therein]{ChaplinMiglio2013, HoudekDupret2015}. Their frequencies are susceptible to the stellar structure, enabling the determination of key parameters such as mass, radius \citep{Kjeldsen1995, Kallinger2010}, and age \citep[e.g.,][]{Creevey2017, Anders2023}. The discovery of mixed modes \citep{Beck2011, Bedding2011} extended the seismic analysis into the central regions of the star, allowing for the determination of the evolutionary state \citep{Bedding2011, Mosser2014} and constraining the internal rotational gradient \citep{Beck2012}. Binary systems play an essential role in seismology for calibrating the scaling relations \citep{Gaulme2016, Benbakoura2021, Beck2025Majestixs} and stellar models \citep{Grossmann2025, Schimak2025, Thomsen2025}.

Thanks to continuous observations from NASA missions such as \Kepler, K2 \citep{Howell2014}, and TESS \citep[Transiting Exoplanet Survey Satellite;][]{Ricker2014}, and ESA’s \Gaia \citep{ESAGaiaPrusti2016}, a large and growing sample of solar-like oscillators in binary systems is now available. In particular, the Non-Single Star catalog \citep[NSS;][]{Arenou2023} from \Gaia Data\,Release\,3 \citep[DR3;][]{GaiaDR3Vallenari2022} has provided orbital solutions for approximately 500,000 systems, along with numerous binary indicators. 
This treasure trove of data offers unprecedented opportunities to investigate the mutual influence of binarity and stellar evolution in exquisite detail. 
By combining data from \Gaia, \Kepler, and TESS, \citet{Beck2022,Beck2024} significantly expanded the sample of known solar-like oscillators in binary systems across all evolutionary stages with detectable oscillations.

In this paper, we investigate the evolution of the binary fraction, orbital properties, and mass distribution of red-giant stars in binary systems by leveraging the combination of asteroseismic constraints and \Gaia DR3 binary solutions. We analyze samples of oscillating red giants from the \apo\ \citep{Pinsonneault2025} and \apok\ \citep{Zinn2022} catalogs, supplemented with homogeneous spectroscopic parameters from APOGEE  \citep[Apache Point Observatory Galactic Evolution Experiment;][]{Majewski2017} (Sect.\,\ref{sec:SampleDefinition}). By cross-matching these samples with the NSS orbital solutions and binary indicators, we study how the binary fraction varies with stellar mass and evolutionary state (Sect.\,\ref{sec:BinaryFraction}), and explore the impact of binarity on the red-giant mass distribution (Sect.\,\ref{sec:MassEvolution}). In Sect.\,\ref{sec:RVevolution}, we examine the attenuation of RV amplitudes, which may indicate common-envelope evolution or merger events. Our findings provide new empirical constraints on the coevolution of stellar and orbital properties in binary systems hosting evolved stars. Conclusions are presented in Sect.\,\ref{sec:DiscussionConclusions}.

\section{Sample definition \label{sec:SampleDefinition}}

\begin{figure*}[t!]
    \centering
    %\vspace{-8mm}
    \includegraphics[width=1.0\textwidth]{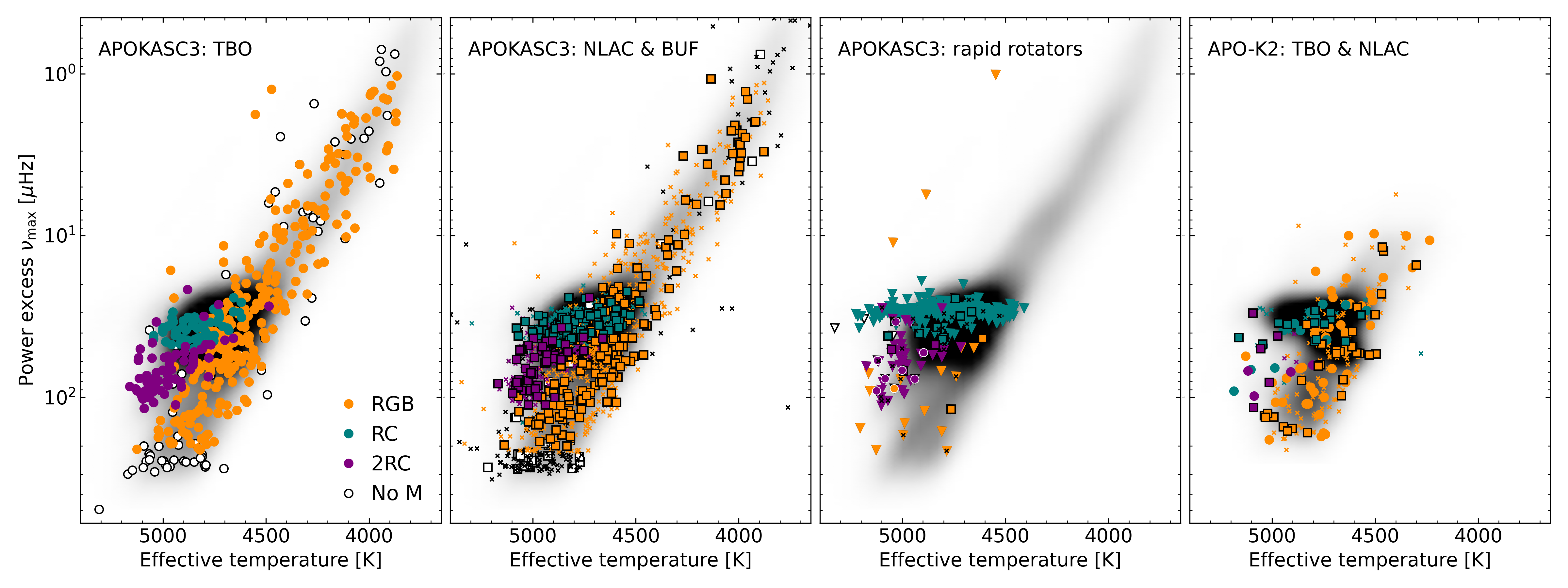}%\vspace{-3mm}
    \caption{Seismic Hertzsprung-Russell diagrams of the investigated catalogs and binary samples. 
   The two panels from left to right depict the targets in the \apo catalog, with solutions in the \Gaia catalogs of Two-Body Orbits (TBOs) as dots and nonlinear or acceleration solutions (NLACs) from NSS as squares. Additionally, the second panel shows the Binary Union Flag (x), which RUWE dominates. The third panel shows the stars with rapid surface rotation ($v\sin i$\,$\geq$5\,km/s) as triangles. Rapidly rotating targets found in binary systems are depicted as dots and squares or marked with crosses, depending on their solution type. The right panel presents TBO, NLAC, and RUWE binary indicators for targets in the \apok catalog. The marker's surface color indicates its seismically determined evolutionary state, as indicated by the legend in the left panel; RGB: red-giant branch, RC: red clump, 2RC: secondary clump, No EvSt: no evolutionary state determined but mass reported, “No M” indicates lower-quality data that did not allow for a seismically inferred stellar mass. Stars with NLAC solutions are further marked with black edges.    
\label{fig:seismicHRD} }%\vspace{-1mm}
\end{figure*}

Following the approach of \citet{Beck2022, Beck2024}, we work under the hypothesis that the oscillating component is the more massive, primary component. The more massive primary evolves faster and is typically the brighter star during the H-shell burning phase of RGB. Because only the mode heights, not their frequencies, are affected by photometric dilution, the presence of a secondary does not influence the determination of global parameters \citep{Beck2018Asterix, Sekaran2019}. When another bright giant is present, the reduced signal-to-noise ratio will substantially increase the parameter uncertainty. 

The assumption that the photometrically dominating star is the primary does not hold for systems where the primary has already reached the He-core burning red-clump (RC) phase, while the secondary still progresses on the RGB, probably more luminous than the primary. In such systems, the mass ratio may significantly deviate from the initial value due to increased mass loss at the RGB tip, and it can even be inverted \citep{Schimak2025}. Given the brief duration of the red-giant phase, such systems are expected to be rare \citep{Miglio2014, Mazzi2025}.

\subsection{\apo catalog}
The \apo catalog\footnote{The \apo catalog files are electronically available for download at \href{https://zenodo.org/records/13308665}{https://zenodo.org/records/13308665}} by \cite{Pinsonneault2025} presents the final and most extensive compilation of the joint survey of asteroseismology of solar-like stars, combining data from the \kepler mission with consistent spectroscopy from APOGEE \citep[for earlier versions, see APOKASC\,1 and 2, in][respectively]{Pinsonneault2014, Pinsonneault2018}.
In its third iteration, the catalog provides mass and radius estimates for more than 12,000 red giants, with the excess of their oscillation power, \num, in the range 300\,$\gtrsim$\,$\nu_\mathrm{max}$\,[$\mu$Hz]\,$\gtrsim$\,0.1 (Fig.\,\ref{fig:seismicHRD}). For each star, values from APOGEE DR17 %\citep[respectively]{Apogee2020Ahumada, Apogee2022Abdurrouf}
 are available, providing a consistent set of fundamental spectroscopic parameters.

Stellar masses were derived from the asteroseismic scaling relations \citep{Brown1991, Kjeldsen1995, Kallinger2010} for 10,036 stars, whose power spectral density allowed for reliable measurements of the global seismic parameters, which are the peak frequency of the oscillation's power excess, \num, and the large-frequency separation between consecutive radial orders, \dnu. The obtained seismic parameters were corrected following the procedure of \cite{Mosser2012a} to account for their departure from the regime of asymptotic oscillation theory. The stellar radii were calculated from the luminosity, using {\Gaia DR3 data \citep[for details, see][]{Zinn2021}.}

\apo presents the seismically inferred evolutionary state of the oscillating red giant based on the dipole-mixed-mode period-spacing measurements (\dpi) presented by \cite{Vrard2024}. This parameter distinguishes the evolutionary phase, specifically RGB and He-core burning stars \citep{Bedding2011, Mosser2014}. However, the catalog does not discriminate between RC stars and more massive secondary-clump (2RC) stars, which both are in the phase of quiescent He-core burning but differ in whether helium ignition occurred under degenerate or non-degenerate conditions. In our analysis, we distinguish them based on their seismically inferred masses, using 1.8\,M$_\odot$ as the threshold. The color code in Fig.\,\ref{fig:seismicHRD} discriminates between these three evolutionary states in the giant phase.

The distinction between RGB and AGB stars becomes increasingly uncertain for more luminous stars than the RC. As is noted by \citet{Pinsonneault2025}, there is approximately a one-in-six chance that a star identified as being on the RGB is, in fact, on the AGB. Seismic discrimination between these two evolutionary phases remains challenging \citep{Dreau2021}. Following \citet{Pinsonneault2025}, we treat AGB stars like RGB stars throughout our analysis. Using seismic parameters, inferred masses and evolutionary states were also used to obtain the stellar ages from comparison with single-star models.

For the 10,036 stars with high-quality asteroseismic solutions, the median fractional uncertainties in \num and \dnu are both 0.6\%. The fractional uncertainties for the mass, radius, and age are 3.8\%, 1.8\%, and 11.1\%, respectively. We adopt these parameters in our analysis as reported in \cite{Pinsonneault2025}.
The catalog includes an additional set of 1,624 stars with lower-quality data, but no asteroseismic solutions are provided for these stars. For completeness, we also searched these systems for indications of binarity.

\subsection{\apok catalog}
\begin{figure}[t!]
    \centering
    \includegraphics[width=1\columnwidth]{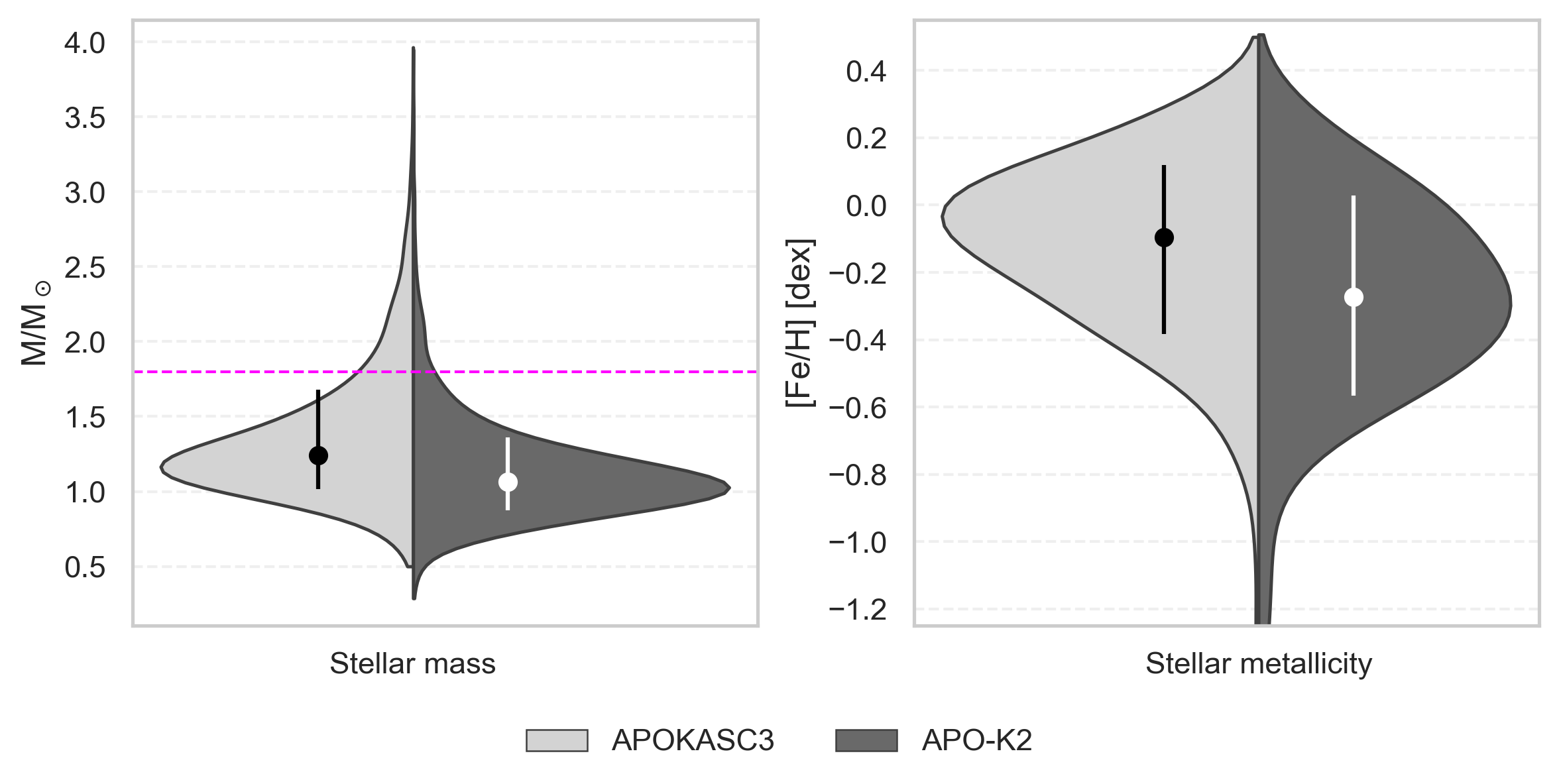} \vspace{-5mm}
    \caption{Normalized kernel density estimates of the distributions for the stellar mass (left) and metallicity (right) for the \apo (light) \apok (dark) samples. The dots and vertical bars mark the respective sample's median and the interquartile range (25$^\mathrm{th}$–75$^\mathrm{th}$ percentiles). The dashed magenta line indicates the mass threshold of 1.8\,M$_\odot$. \label{fig:ApoMasses} }
    \vspace{-3mm}
\end{figure}

The third version of the APO-K2 Catalog\footnote{The APO-K2 catalog files are electronically available for download from CDS through the reference \href{https://cdsarc.cds.unistra.fr/viz-bin/cat/J/ApJ/926/191}{J/ApJ/926/191}.} by \cite{Zinn2022} presents a similarly comprehensive study, using APOGEE spectroscopy and asteroseismic detections of 7,673 oscillating red-giant stars from the Campaigns C1-C8 and C10-18 of the K2 mission. As this work predates \Gaia DR3, it uses DR2 parallaxes and a different correction method for \dnu. Furthermore,
%photometric temperatures and metallicities are used. 
spectroscopic values from the APOGEE DR16 and GALAH DR2 are used.
Due to the shorter time base and observational systematics of the K2 mission introduced by the satellite repositioning every $\sim$6\,hours, stars in the \apok sample are found in a smaller range of oscillation frequencies, 200\,$\gtrsim$\,$\nu_\mathrm{max}$\,[$\mu$Hz]\,$\gtrsim$\,5 (Fig.\,\ref{fig:seismicHRD}). 
For this sample, the median fractional uncertainties in \num and \dnu are better than 2\% and 1\%, respectively. The fractional uncertainties for mass, radius, and age are $\sim$11\%, $\sim$7\%, and $\sim$20\%, respectively.

The two samples appear to be significantly shifted when comparing the normalized kernel density estimates (KDE) of the stellar masses (Fig.\,\ref{fig:ApoMasses}), reported in the \apo and the \apok catalogs. However, the differences between the two works mentioned above amount to only slight variations. While the difference in parallax scale between DR2 and DR3  as well as the different correction of the scaling relations may introduce the most considerable shift, we argue that the leading cause is that K2 and \Kepler sample stellar populations of different metallicities (Fig.\,\ref{fig:ApoMasses}), as the two missions probed different galactic latitudes (Fig.\,\ref{fig:galacticPlane}).
A comparison by \cite{Warfield2024} of the ages reported in the two samples found good agreement when binned by distance to the Galactic plane. This finding suggests that this difference is due to a population effect but does not rule out a small systematic difference. Therefore, we treat the two samples separately.

\subsection{Binary sample selection}
As a source of binarity indicators, we used the non-single-star catalog \citep[NSS\change{,}][]{Arenou2023}, published as part of the third data release (DR3) of the ESA \Gaia mission \citep{ESAGaiaPrusti2016, GaiaDR3Vallenari2022}. It provides several sub-catalogs.  
The Gaia Collaboration cautions that the selection effects of the NSS catalog are difficult to quantify due to the heterogeneous detection algorithms used in its processing pipeline \citep{Halbwachs2023}. Recent studies have started to address this challenge by modeling the detection probabilities of astrometric binaries \citep{ElBadry2024, Lam2024}.

First, we used the catalog of two-body orbit solutions (TBO), which presents about half a million orbital parameters derived from the RV time series, obtained with the {\Gaia Radial Velocity Spectrometer} \citep[RVS,][]{Katz2022}, astrometry \citep{Holl2023, Halbwachs2023} and photometry \citep{Mowlavi2023}. \cite{Arenou2023} point out that these are binary candidates for which completeness was chosen over purity. We followed a similar approach and did not filter solutions based on \GaiaDR significance.
From a detailed comparison of the SB9 catalog   \citep[Ninth Catalog of Spectroscopic Binary Orbits;][]{Pourbaix2004}, \cite{Beck2024} showed a good agreement for orbital periods P$_\mathrm{orb}$\,$\lesssim$\,1,500\,days. 
We adopt a solution from the TBO as a binary candidate if at least the orbital period is provided in the catalog. For stars for which a single star model cannot sufficiently explain the proper motion and no spectroscopic solution for an orbit was found, \cite{Arenou2023} provides nonlinear or acceleration solutions (hereafter NLAC). The comparison with the SB9 confirmed that NLAC solutions indicate binaries with orbital periods typically much longer than 1,000\,days \citep{Beck2024}. 

An additional binarity indicator from \Gaia is the renormalized unit weight (RUWE). This quantity expresses the quality of the astrometric solution for a single star model. RUWE values larger than 1.4 are considered good indicators of unresolved binaries \citep{Belokurov2020RUWE}. However, even among the large sample of the SB9, $\sim$40\% of all confirmed binary systems, in particular distant but short periodic systems, show RUWE\,$\lesssim$\,1.4 \citep{Beck2024}. RUWE, therefore, is an indicator for possible binarity but is not suited as a diagnostic to exclude a secondary companion. In their comprehensive catalog\footnote{The \Kepler-\Gaia catalog files are electronically available for download from \href{https://zenodo.org/records/14774100}{https://zenodo.org/records/14774100}.}, \cite{GodoyRivera2025} have conducted a detailed search amending the binary indicators of \Gaia by membership in binary catalogs from ground-based observations for \Kepler targets, presented in the Binary Union Flag (BUF). Given the catalog's constructive nature, this binary flag is dominated by RUWE as the primary indicator.

In total, we find 1296 binary candidates that are brighter than G\,$\leq$\,13\,mag. Table\,\ref{tab:binSamples} provides the individual sample sizes. 
The position of stars from the \apo and \apok samples that are listed in the NSS are depicted in the first, second, and fourth panels of Fig.\,\ref{fig:seismicHRD}. 
The parameters for these systems from the \apo and \apok samples are available in the electronic Tables\,\ref{tab:A1} and \ref{tab:A2}, respectively, with their inventory explained in detail in App.\,\ref{App:A}.
Additionally, we added 49\,binaries with oscillating MS or SG primaries\footnote{The \Kepler/TESS-\Gaia Binary catalog files are electronically available for download from CDS through the reference \href{https://cdsarc.cds.unistra.fr/viz-bin/cat/J/A+A/682/A7}{J/A+A/682/A7}.} from \cite{Beck2024}.

\subsection{Rapidly rotating red giants}
Finally, we focus on the rapid rotators subsample in the \apo sample. In a previous study, \cite{Patton2024} has identified $\sim$1500 active red-giant candidates from the full APOGEE\,DR16, whose projected rotation velocity exceeds $v\sin i$\,$\geq$\,5\,km/s.  For the stars with asteroseismic solutions from \Kepler observations, \cite{Pinsonneault2025} determined mass, radius, and the evolutionary state for 332 objects. We include this subsample, depicted in Fig.\,\ref{fig:ApoMasses}, as enhanced rotation could be the signature of merger products or post-common envelope phases \citep{Kochanek2014}. Table\,\ref{tab:rapidRotBinaries} provides the parameters of the 14 systems that host rapidly rotating giants.

\section{Evolution of binary fraction \label{sec:BinaryFraction}}
Systems with orbital periods that are not sufficiently wide to allow the giant to expand freely until the tip of the RGB will undergo a common-envelope (CE) phase, likely leading to a short-lived destructive interaction. Consequently, the binary fraction is expected to decrease between the MS and the RGB and RC phases. However, the lack of consistent and complete binary indicators for large ensembles in the past made it difficult to estimate the evolutionary depletion of binary systems.
From their large-sample study of RV variations in APOGEE data \cite{Badenes2018}, they reported a $\sim$60\% reduction from MS to RGB, with an estimated uncertainty of $\sim$20\%.

The determination of the binary fraction relies on stellar multiplicity indicators rather than on the full set of orbital parameters for each system. Therefore, we used the  BUF from \citet[see Table\,1]{GodoyRivera2025}, which offers the most comprehensive compilation of all available binary indicators for stars in the \Kepler field of view, from \Gaia with
RUWE  (33.8\%),
RV Variable (5.7\%),
NSS  (5.6\%),
Eclipsing Kepler  (4.0\%),
Eclipsing Gaia  (1.2\%),
as well as entries from ground-based surveys, like the SB9 (0.1\%),
binary stars from the NASA exoplanet archive (0.1\%),
indications from the HIPPARCOS-Gaia catalog of accelerations
(HGCA) (0.1\%), and entries from the
Washington double star (WDS) catalog (4.0\%). We excluded the \Gaia Variablity flag ($\sim$1.1\%).
An additional 44.2\% of stars were found in more than one of these categories.
As such, compilation of binary markers is not yet available for the \apok sample, we only compute the binary fraction from the \apo sample, as combining both samples would lead to a skewed result due to the differing completeness factors between the two catalogs. 

The rich, homogeneous asteroseismic and spectroscopic dataset of the \apo catalog provides an ideal opportunity to investigate the reduction in the binary fraction with advancing stellar evolution. 
We constructed a magnitude-limited sample to determine a robust estimate of the binary fraction. The binary detection using \gaia shows an explicit cutoff with G\,$\lesssim$\,13\,mag \citep[][]{Arenou2023}, which is brighter than the faint and well-populated end of the sample of oscillating red giants from the \kepler mission (G\,$\lesssim$\,15\,mag). We therefore adopted $G$\,$<$\,13\,mag as the faint and \textit{Gaia}’s saturation limit of $G$\,$>$\,4\,mag as the bright cutoff of our sample. 
Table\,\ref{tab:binSamples} reports the number of oscillating stars categorized by the seismically inferred evolutionary state of the asteroseismic primary.

\subsection{Evolutionary subsample definition} 

Depending on its mass, a red-giant star can undergo two very distinct evolutionary channels \citep[for details see the monographs by, e.g.,][]{Kippenhahn2013, Salaris2005, Pinsonneault2023}. The masses of giants in the \apo catalog are dominated by low-mass stars but extend up to  $\sim$3\,M$_\odot$ (Fig.\,\ref{fig:ApoMasses}).
If the stellar mass is M/M$_\odot$\,$\lesssim$\,1.8, the inert helium core degenerates fully before reaching the ignition temperature of He. As a result, the star must ascend the RGB and continue to expand while the core heats up. With the onset of quiescent He-core burning at the tip of the RGB (up to R/R$_\odot$\,$\simeq$\,200\,$\simeq$\,1\,AU), following a series of off-center helium flashes, the star settles into the RC phase \citep{bildsten2012}. 
Stars more massive than this limit (M/M$_\odot$\,$>$\,1.8) ignite their helium core before it degenerates, doing so at significantly smaller radii on the RGB (200\,$\gtrsim$\,R/R$_\odot$\,$\gtrsim$\,80), and subsequently settle into the 2RC. 

Furthermore, we must consider the differing binary fractions of progenitors of the red-giant stars in the sample. On the MS, the binary fraction is a strong function of the mass of the primary component \citep{Lada2006}. The main-sequence progenitors of these more massive stars are early to mid-A-type stars. While 40\,-\,50\% of all solar-like dwarf stars (G and mid F) are found to be in binary systems, this fraction increases to 50\,-\,70\% for the more massive early F- and A-type stars on the MS \citep[e.g.,][]{Offner2022, MoeStefano2017}. 
   
The precisely seismically inferred masses allow for an effective separation between the degenerate and non-degenerate evolutionary channels in the red-giant phase, enabling tests for these population differences. We therefore evaluated the binary fraction separately for these two evolutionary regimes (M\,$\gtrless$\,1.8\,M$_\odot$).

\subsection{Binary fraction evolution for solar-like stars into the RC}

Solar-like oscillating dwarfs constitute the progenitor sample for the giant stars that follow the evolutionary channel leading into degenerate He-core ignition. We therefore used the sample of 620 oscillating solar-like MS and subgiant (SG) stars observed by \Kepler \citep[][]{Chaplin2011,Mathur2022} to estimate the MS baseline binary fraction.

Using the Union Binary Flag from \cite{GodoyRivera2025}, we identify 195 binary candidates in this sample, yielding a binary fraction of 31.5\%. This value is in good agreement with the binary fraction of 35\% reported by \cite{Badenes2018} for MS stars in the sample of APOGEE, showing that for this sample, the possible selection bias potentially introduced by basing the analysis on solar-like oscillating stars is small for the reference sample.
However, both values are well below the expected $\sim$50\%, suggested by large-sample statistics. For \cite{Badenes2018}, this discrepancy can be explained by the nature of APOGEE spectroscopy, where some binaries remain undetected due to the suboptimal sampling of the observations or inclination effects. The binary indicators in the Union Binary Flag are dominated by \Gaia RUWE measurements (accounting for 76.5\% of binary candidate indicators). 
As mentioned before, $\sim$40\% of confirmed SB9 binaries have RUWE\,$\leq$\,1.4 \citep{Beck2024}, which suggests that a significant fraction of binaries in our sample might remain \hbox{undetected due to weak photocentric motion}.

For the computation of the binary rate on the MS, we should consider that this sample could be biased against binarity. From the sample of \cite{Mathur2022} we find that about 50\% of all dwarfs observed by \Kepler in the one-minute (short cadence) sampling show solar-like oscillations. \cite{Mathur2019} argued that the non-detection of mode is probably caused by high levels of activity, low metallicity, or tidal interaction, which reduce the oscillation amplitudes below the detection limit. The latter assumption was motivated by the finding that red giants in binary systems show suppressed oscillation, once the fractional radius of both components fills more than $\sim$20\% of the semimajor axis, a, $(R_1+R_2)/a$\,$\gtrsim$\,20\%, and the tidal spin up activates a stellar dynamo \citep{Gaulme2014}. Such a fractional radius coincides with the threshold of tidal interaction through the equilibrium tide for the circularization and synchronization \citep{Beck2018Tides}. However, \cite{Beck2024} showed that solar-like stars in strongly interacting binary systems that are synchronized and circularized by the dynamical tide \citep[$P_\mathrm{orb}$\,$\lesssim$\,8\,days,][]{Zahn1989, Raghavan2010} can exhibit solar-like oscillations even at high levels of photospheric activity. Among the 663 MS and SG stars that were observed in short cadence and did not show a significant excess of oscillation, we find 270 binary candidates from the binary union flag, which corresponds to a binary fraction of $\sim$40.7\%, which suggests that binarity might have a small effect but is not the dominant mechanism to reduce the amplitude of solar-like oscillations in MS stars. Therefore, we consider this base value provides a good reference for the binary fraction on the MS. 

The binary fractions reported in this work are assumed to be internally consistent within the seismic \textit{Kepler} samples.
However, both the detection of solar-like oscillations and the identification of binaries in \textit{Gaia} are subject to complex and non-uniform selection effects.
The completeness of the seismic sample depends on \textit{Kepler}'s target selection, apparent magnitude, and local crowding conditions \citep{Wolniewicz2021}, while the detectability of binaries in \textit{Gaia} varies with magnitude, orbital period, and scanning-law coverage \citep{Cantat-Gaudin2023, Castro-Ginard2023}.
RUWE, the primary indicator used to identify binaries in this study, is known to miss a significant fraction of confirmed binaries and may also yield false positives \citep{Cantat-Gaudin2023, Beck2024}. As such, the inferred binary fractions should be interpreted as reflecting relative trends across different evolutionary stages, rather than representing an absolute census of binaries in the field population. 
In light of these limitations, it is beyond the scope of this paper to generalize the reported binary rates or to estimate the absolute binary fraction and its associated uncertainties for the broader Galactic population. A full forward-modeling of the joint \textit{Kepler}--\textit{Gaia} selection function is necessary for such an analysis and is deferred to future work.

 \begin{table*}[t!]
    \centering
\tabcolsep=9pt
\caption{Binary fraction in the sub-phases of the red-giant star evolution.}% \vspace{-2mm}
    \begin{tabular}{lr|rrr|rrr}
\hline\hline %& 
\multicolumn{1}{c}{Evolutionary state}
 & \multicolumn{1}{c}{Full Sample} 
 & \multicolumn{3}{c}{Mag.\,Limited (M/M$_\odot$\,$\leq$ 1.8)} 
 & \multicolumn{3}{c}{Mag.\,Limited (M/M$_\odot$\,$>$ 1.8)}   \\
 & \multicolumn{1}{c}{Stars}
 & \multicolumn{1}{c}{Stars} & \multicolumn{1}{c}{Bin.Sys.} & \multicolumn{1}{c}{Bin.Frac.} 
 & \multicolumn{1}{c}{Stars} & \multicolumn{1}{c}{Bin.Sys.} & \multicolumn{1}{c}{Bin.Frac.} \\[1mm] \hline

MS \& SG (oscillation detected) & 620        
	 & 620 & 195 & 31.5 \%        
	 & $-$ & $-$ & $-$ \\
     
MS \& SG (no oscillation) & 663        
	 & 663 & 270 & 40.7 \%        
	 & $-$ & $-$ & $-$ \\[1mm] \hline

H-shell burning \change{(}RGB\change{)}  & 7191        
	 & 5331 & 438 & 8.2 \%        
	 &  447 & 43 & 9.6 \%  \\[1mm] 
\multicolumn{1}{r}{low-luminosity RGB}  & 4103        
	 & 3209 & 310 & 9.7 \%        
	 &  139 & 18 & 12.9 \% \\ 
\multicolumn{1}{r}{high-luminosity RGB}  & 3088        
	 & 2122 & 128 & 6.0 \%        
	 &  308 & 25 & 8.1 \% \\[1mm] \hline 

He-core buring (RC \& 2RC) & 5560        
	 & 3881 & 200 & 5.2 \%        
	 & 982 & 134 & 13.6 \% \\[1mm] \hline
	 	 	 	 
\end{tabular}
\tablefoot{The left panel specifies the evolutionary states and gives the number of all stars with this identified phase in the catalog. The red-giant stars are based on \apo \citep{Pinsonneault2025}, while the MS dwarfs and subgiants are taken from the sample of \cite{Mathur2022}. For the RGB, we first provide the full range and then separate between the case of the low- ($\nu_\mathrm{max}$\,$\geq$\,30\,$\mu$Hz) and high-luminosity regime ($\nu_\mathrm{max}$\,$<$\,30\,$\mu$Hz).
The center and right panel separate the stars by their mass regimes leading to the degenerated (M/M$_\odot$\,$\leq$ 1.8) and non-degenerated (M/M$_\odot$\,$>$ 1.8) ignition of the He core, respectively.
Both panels' first, second, and third columns indicate the number of stars, binaries identified through the Binary\,Union\,Flag  \citep{GodoyRivera2025}, and the corresponding binary fraction in percent for the respective evolutionary state.
\vspace{-2mm}}
    \label{tab:binSamples}
\end{table*}

\begin{figure}[t!]
    \centering
    \vspace{-2mm}
    \includegraphics[width=\columnwidth]{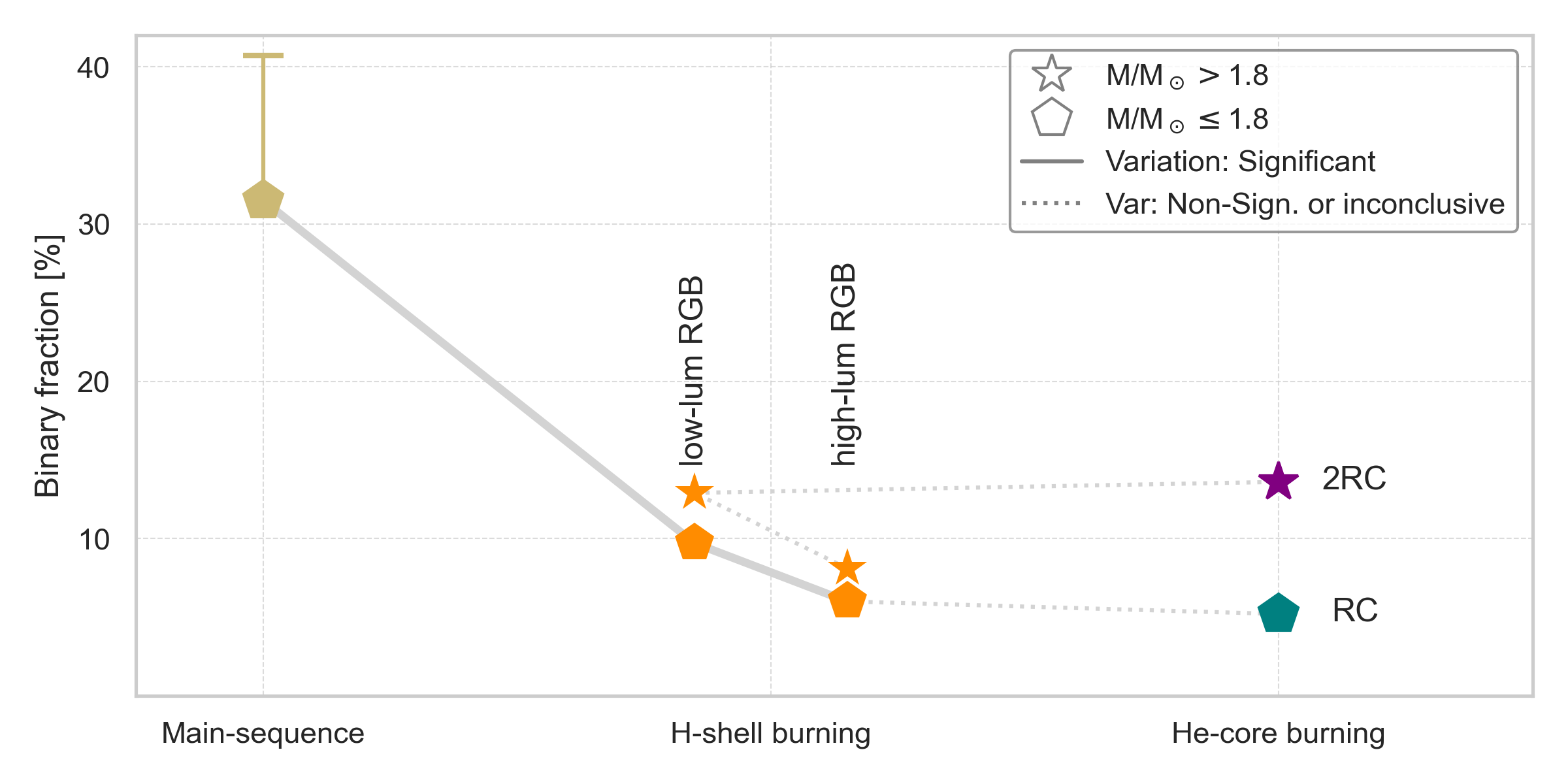}\vspace{-2mm}
    \caption{Binary fraction as a function of their evolutionary states. Pentagons mark stars that undergo core-degeneration on the RGB (M/M$_\odot$\,$\leq$\,1.8), while star symbols indicate the binary fraction for stars that will ignite He under non-degenerate conditions  (M/M$_\odot$\,$>$\,1.8).  The style of the connecting lines depicts the significance of the variations as discussed in the text and specified in Table\,\ref{tab:binReductionRate}. The yellow bar indicates the binary fraction among MS stars without oscillations. The marker color depicts the evolutionary state as indicated on the horizontal axis. \vspace{-5mm}
 \label{fig:binaryFraction} }\end{figure}

For stars with masses below 1.8\,M$_\odot$, we find 438 binary candidates with RGB primaries and 200 with primaries in the RC, corresponding to binary fractions of 8.2\% for the H-shell and 5.2\% for the He-core burning stars that exhibit solar-like oscillations (Table\,\ref{fig:binaryFraction} and Fig.\,\ref{tab:binSamples}). From these numbers, we estimate a binary attrition from the MS to the RGB phase to be $\sim$69\% (Table\,\ref{tab:binReductionRate}), which is in good agreement with the reduction found in the APOGEE sample \citep{Badenes2018}.  The binary depletion is even larger between the MS and the RC phase with $\sim$84\%. 

With a maximum change of the stellar radius of nearly two orders of magnitude, the RGB is the phase with the most significant variations and differences in stellar radius. Therefore, a significant reduction along the RGB evolution itself can be expected. To test this variation, we used the peak frequency of the oscillatory power excess, \num, to split the RGB into two subsamples and evaluate their binary fractions separately. 
For the low-luminosity RGB (300\,$\gtrsim$\,$\nu_\mathrm{max}$\,$\geq$\,30\,$\mu$Hz, which correspond to approximately to a range in radius of 3\,$\lesssim$\,R/R$_\odot$\,$\lesssim$\,10) we find a binary fraction of 9.7\%. For the high-luminosity RGB (30\,$>$\,$\nu_\mathrm{max}$\,$\gtrsim$\,0.1\,$\mu$Hz, 10\,$\lesssim$\,R/R$_\odot$\,$\lesssim$\,100) we find a binary fraction of 6.0\%. This corresponds to an attrition of $\sim$38\% from the low- to the high-luminosity RGB, of $\sim$81\% with respect to the MS. For the detailed overview of the values, we refer the reader to Fig.\,\ref{fig:binaryFraction}, as well as Tables\,\ref{tab:binSamples} and \ref{tab:binReductionRate}. 

\begin{table}[t!]
    \centering
\tabcolsep=6pt
\caption{Binary attrition between evolutionary states and samples.}%\vspace{-2mm}
    \begin{tabular}{lcc}
\hline\hline %& \multicolumn{1}{c}{\apok}
\multicolumn{1}{c}{Track or comparison} 
& \multicolumn{1}{c}{Rel.\,Decrease}
& \multicolumn{1}{c}{$p$} \\ 
[1mm] \hline
MS\,$\to$\,low-lum. RGB                 & 69.3 \%    & 10$^{-40~}$  (h)\\
MS\,$\to$\,high-lum. RGB                & 80.8 \%    & 10$^{-66~}$ (h)  \\
MS\,$\to$\,RC                           & 83.6 \%    & 10$^{-102}$ (h)  \\[1mm]
low-lum. RGB\,$\to$\,high-lum. RGB      & 37.6 \%    & 10$^{-6~~}$ (h) \\
low-lum. RGB\,$\to$\,RC                 & 46.7 \%    & 10$^{-13~}$ (h)  \\
high-lum. RGB\,$\to$\,RC                & 14.6 \%  & 0.15 (n)\\[1mm] \hline	 
low-lum. RGB\,$\to$\,high-lum. RGB      & $-$   & 0.12 (n) \\
high-lum. RGB\,$\to$\,2RC               & $-$   & 0.93 (n)  \\[1mm] \hline	
osc. MS\,$\Leftrightarrow$\,non-osc MS  & 29.2 \% & 10$^{-4}$ (h) \\
deg RGB\,$\Leftrightarrow$\,non-deg RGB & 17.1 \% & 0.01 (m) \\[1mm] \hline	
\end{tabular}
    \tablefoot{The table reports the significance levels of the binary attrition depicted in Fig.\,\ref{fig:binaryFraction}. The first column indicates the evolutionary states that are compared. The second column indicates the relative decrease in the binary fraction, expressed as a percentage. The final column indicates the significance of the variation from a $\chi^2$ test, as described in the text. The table's top and middle panels show the tracks that undergo He-core ignition under degenerate and non-degenerate conditions. The bottom panel compares the average binary fraction between the indicated samples. } \vspace{-3mm}
    \label{tab:binReductionRate} 
\end{table}

We applied the $\chi^2$ test for independence to evaluate whether the observed differences in binary fractions between evolutionary states are statistically significant, under the null hypothesis that binarity is independent of stellar evolutionary phase. We adopt the traditional threshold-based classification of $p$ values, whereby results with $p$\,$<$\,0.001 are deemed highly significant, indicating strong statistical evidence against the null hypothesis. When $p$ values fall in the range 0.001\,$\leq$\,$p$\,$<$\,0.05, the variation is deemed marginally significant, suggesting a moderate likelihood that the observed difference is not purely due to statistical fluctuations. Conversely, if $p$\,$\geq$\,0.05, we classify the variation as insignificant. 
The significance test, whose results are reported in Table\,\ref{tab:binReductionRate} and depicted in Fig.\,\ref{fig:binaryFraction}, shows that the binary attrition from the MS to any state of the giant branch and the RGB to the RC are significant. It is worth highlighting that the $\sim$38\% reduction between the low- and high-luminosity regimes of the RGB evolution is highly significant. 
The only variation not found to be significant in the low-mass sample is the reduction of $\sim$15\% from the high-luminosity RGB into the RC phase. To determine whether the small variation between the high-luminosity RGB and the RC is real, a larger sample would be required. 

Finally, we note that two important potential biases exist for this analysis. First, the selection criterion on solar-like oscillations introduces a bias. As mentioned before, it was shown that solar-like oscillations are suppressed in giants where the combined fractional stellar radii exceed $\sim$20\% of the system's semimajor axis.
Consequently, stars starting to fill their Roche-lobe substantially will also be excluded from the sample of oscillating giants. However, once this phase has started, the timescales of a few hundred\,kiloyears, leading into a potentially destructive evolution and white-dwarf phase, are short. Given the duration of the RGB phase, this dilution effect does not significantly skew the sample.  Second, stellar oscillations are also detectable from RV variations. However, only on the super-luminous RGB ($\nu_\mathrm{max}$\,$\lesssim$\,0.1\,$\mu$Hz, or 100\,$\lesssim$\,R/R$_\odot$\,$\lesssim$\,200), the RV amplitude of stellar oscillations becomes substantial enough to reach or exceed the RV threshold of $\sim$1.4\,km/s \citep{Katz2019} for faint stars of the \Gaia RVS instrument ($G$\,$\lesssim$\,12\,mag). As this regime is not probed by the \num range in \apo, possible false positive binary solutions produced by intrinsic stellar RV variations do not bias this comparison. We therefore assumed the completeness constant throughout the studied evolutionary states. 

\begin{figure*}[t!]
    \centering
    %\vspace{-8mm}
    \includegraphics[width=0.85\textwidth]{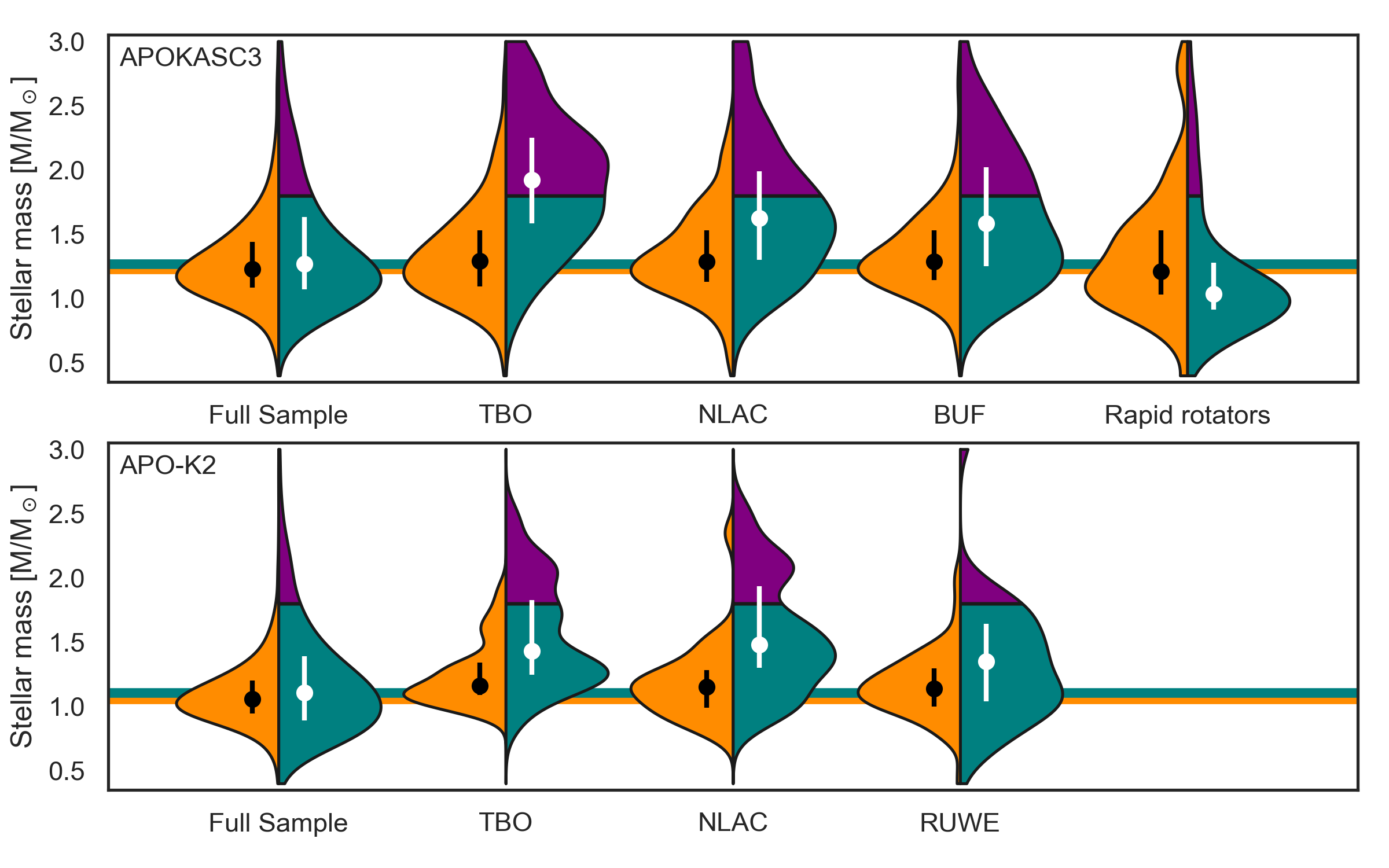}\vspace{-3mm}
    \caption{
Normalized kernel density estimates of the stellar mass distributions for red-giant stars in the \apo (top) and \apok (bottom) samples. Each KDE is split by evolutionary stage, with the left side (orange) showing RGB stars and the right side displaying core-helium burning stars. The latter are further divided into RC (teal) and 2RC (purple) stars, separated at 1.8\,M$_\odot$. Distributions are shown for the full sample and several binary candidate subsamples identified by the TBO, NLAC, and Binary Union Flag (BUF) for \apo and RUWE for \apok, as well as the rapid rotator subset in \apo. Dots and vertical bars indicate the median and interquartile range (25$^\mathrm{th}$–75$^\mathrm{th}$ percentiles) for each population. Horizontal orange and teal lines mark the median RGB and RC stellar masses of the respective full samples.
\label{fig:massHistograms} 
}
\vspace{-3mm}
\end{figure*}

\subsection{Evolution of the binary fraction for 2RC giants}

The progenitors of secondary clump stars are early F- and A-type stars, which exhibit a different type of pulsations \citep[$\gamma$\,Dor and $\delta$\,Scuti,][and references therein]{Aerts2010}. Therefore, these are not represented in the MS sample of \cite{Mathur2022}. We measure the binary fraction for the non-degenerate evolution starting from the low-luminosity RGB. 

Table\,\ref{tab:binSamples} and Fig.\,\ref{fig:ApoMasses} show that such relatively massive stars represent only a small fraction of the full \apo sample. Consequently, their population and binary counts suffer from small-sample statistics, and all differences found between the subsamples were deemed to be insignificant. The only marginally significant difference that we found from the numbers in Table\,\ref{tab:binReductionRate} was that the binary fraction of the full sample of non-degenerate RGB stars ($\sim$9.6\%) is $\sim$17\% richer in binaries than the full sample of degenerate RGB ($\sim$8.2\%). This increased richness aligns with the differences in the binary fraction of their MS progenitors, as observed in large binary statistics \citep{MoeStefano2017,Offner2022}.

\section{Coevolution of the primary mass and the orbit \label{sec:MassEvolution}}
The key parameter determining the evolution of a star is its initial mass \citep[][and references therein]{Serenelli2021}. It defines the stellar structure and timescales of the evolution. 
In a binary system, the total mass of the components in solar units ($M_\odot$), together with the semimajor axis, $a$ (in AU) or orbital period $P_\mathrm{orb}$ in years, defines the system's configuration through the third law of \cite{Kepler1619},
\begin{equation}    \frac{a^3}{P_\mathrm{orb}^2} \simeq M_1 + M_2.     
%\frac{a^3}{P^2}=\frac{G}{4\pi^2} \cdot M_1 + M_2.
\label{eq:Kepler3}
\end{equation}
Thus, stellar mass dictates the evolution of individual stars and influences the properties and evolution of binary and planetary systems through tidal interactions and mass loss.

Tidal interactions gain importance when the radius of one of the stellar components grows to a significant fraction of the semimajor axis. The redistribution of angular momentum and dissipation of kinetic and potential energy from tides into heat drives the changes in the orbits and the spins. This process leads the system toward a state of minimum energy, where the orbit is circularized, and the stellar spins become aligned and synchronized \citep[][]{Zahn1977, OgilvieLin2007, Mathis2015}. Tidal interaction occurs in waves as a dynamical tide or as a hydrostatic adjustment in the equilibrium tide, whereby the stellar structure, which depends primarily on the stellar mass, determines the dominant tidal mechanism. The various manifestations of tides are intensively studied through the distributions of orbital parameters in large samples from open clusters or from the \gaia mission, providing valuable constraints on tidal dissipation and orbital evolution \citep[e.g.,][]{Mirouh2023, DewberryWu2025}.  

Mass loss during stellar evolution, such as stellar winds, also affects orbital evolution by removing mass and angular momentum from the system \citep{Miglio2012}. Following Kepler's third law (Eq.\,\ref{eq:Kepler3}), this leads to an increase in the orbital period \citep{ZB1989, Soberman1997} and also to the spin-down of the mass-loosing star \citep[e.g.,][]{Skumanich1972}.%, Garcia2014b}. 

The evolution of a binary system becomes particularly complex once one component fills its Roche lobe and significant mass transfer occurs. This process can either shrink or widen the orbital separation, depending on the mass ratio and angular momentum exchange \citep[][and references therein]{Ivanova2013}. At the same time, the donor star's radius is increasing. Red giants, which possess convective envelopes, expand in response to mass loss to preserve an approximately constant entropy profile \citep{Hjellming1987}. Moreover, tidal heating in close systems may further inject energy into the stellar envelope, contributing to additional expansion \citep{Gallet2017}. This expansion, occurring on timescales much shorter than the evolutionary pace, can destabilize mass transfer in close binaries, leading to runaway expansion of the donor.

If the radius of one of the binary components exceeds the orbit of the companion, the system will enter the short-lived, most likely destructive common-envelope (CE) phase, which would end the red-giant phase of the star abruptly with the CE ejection, exposing the “naked” He-core as a white dwarf \citep[][]{Ivanova2013}. Due to the hydrodynamical complexity and their short timescales, this phase is difficult to model and poorly understood \citep{Han2002}. However, several parameters are identified to play a significant role in determining the fate of the binary system. The primary determining factor is the semimajor axis relative to the maximum stellar radius of the components, as this governs both tidal interaction strength and the Roche-lobe radius. Furthermore, the mass ratio between the two stars, which influences the difference in their evolutionary timescales \citep[e.g.,][]{Miglio2014, Beck2018Asterix, Grossmann2025}

Another critical parameter is whether the mass loss is stable or unstable \citep{Podsiadlowski2001}. It is important to note that the mass loss rate of the donor is substantially larger than the accretion rate of the accretor. In case of stable mass loss, the mass ratio $q$ will be gradually adjusted, allowing the system to adapt continuously by expanding its orbit and shifting to longer orbital periods. Such a system will not undergo a common envelope or merger phase. In the case of unstable mass loss, the orbit shrinks, leading to a CE phase. 
It is still unclear what determines the fate of a binary system at the end of the CE phase. If a system enters a CE phase, envelope ejection depends on whether the orbital energy released during the spiral-in phase exceeds the envelope's binding energy. Generally, systems with extreme mass ratios ($q$\,$=$\,$M_2 / M_1$\,$\ll$\,1) are unlikely to inject sufficient orbital energy to expel the envelope, leading to a merger event \citep{Ivanova2013}. 

Stellar ages are typically inferred from model fitting, based on single-star evolutionary models assuming steady mass loss \citep[e.g.,][]{Pinsonneault2025}. Episodic or interaction-induced mass loss in binary systems will significantly over- or underestimate the ages of the post-interaction stellar products.

\subsection{Mass distributions separated by evolutionary state}
To isolate evolutionary effects in the binary sample, we constructed reference mass distributions for the entire sample, separated by evolutionary state. Since the mass distributions differ between the \apo and \apok catalogs, we analyze these two catalogs separately to avoid potential systematic biases in the samples skewing the result. 
The resulting KDE, shown in  Fig.\,\ref{fig:massHistograms}, shows that the mass distribution among the full samples is similar. Furthermore, Fig.\,\ref{fig:massHistograms}  presents KDEs for the mass distributions for samples selected from various binary indicators. In the RGB phase, the mass distribution of all binary markers closely resembles that of the full sample, with only minor deviations.

A very different picture emerges when examining the binary sample of RC and 2RC stars. The mass distribution of the binary primaries that have ignited their He-core is significantly shifted to higher masses (Fig.\,\ref{fig:massHistograms}).
While the full sample of He-core burning stars in the \apo sample has a median mass of 1.27\,M$_\odot$ and the interquartile range (25$^\mathrm{th}$–75$^\mathrm{th}$ percentiles) between 1.07 and 1.63\,M$_\odot$, the median mass for He-core burning primaries of binary systems with orbital periods from the TBO is 1.92\,M$_\odot$ and their interquartile range between 1.58 and 2.25\,M$_\odot$. 
A similar trend is found for binary stars hosting He-core burning stars from the \apok sample and TBO orbital solutions (Fig.\,\ref{fig:massHistograms}). 

We suggest that the upward shift of the median mass in the binary sample is caused by the removal of lower-mass systems that arise from disruptive and merging events along the RGB, as shorter-period binaries undergo a CE phase, mostly leading to merger or envelope stripping of the RGB primary, before such stars can reach the tip of the giant branch and continue into the RC phase. This mechanism preferentially removes systems with $M$\,$\lesssim$\,$1.8$\,$\mathrm{M}_\odot$ and $P_\mathrm{orb}$\,$\lesssim$\,$1000$\,days and 
introduces a stronger mass dependence on the binary fraction than for the MS progenitors. Because the orbital period is the main parameter determining whether a system undergoes the CE phase, we expect a mass–period relation in the He-core-burning phase.

Figure\,\ref{fig:massHistograms} presents similar shifts in the mass distributions of He-core burning stars from samples identified through different binary indicators. 
The most pronounced difference is observed for binaries identified in the \gaia TBO catalog. The comparison with the SB9 catalog by \cite{Beck2024} suggests that the TBO predominantly contains orbits with resolved periods of $P$\,$\lesssim$\,1500 days, while NLAC solutions typically indicate longer-period systems. We argue that the \gaia TBO catalog is more sensitive to the depletion of short-period systems. Since the Binary Union Flag indicator depends largely on RUWE, they are also sensitive to long-period binaries. This trend is consistently observed in both the \apo and \apok catalogs, excluding that this results from a selection bias.

\subsection{Mass dependency of the orbital parameters}
\begin{figure*}[t!]
    \centering
    %\vspace{-8mm}
    \includegraphics[width=1.25\columnwidth]{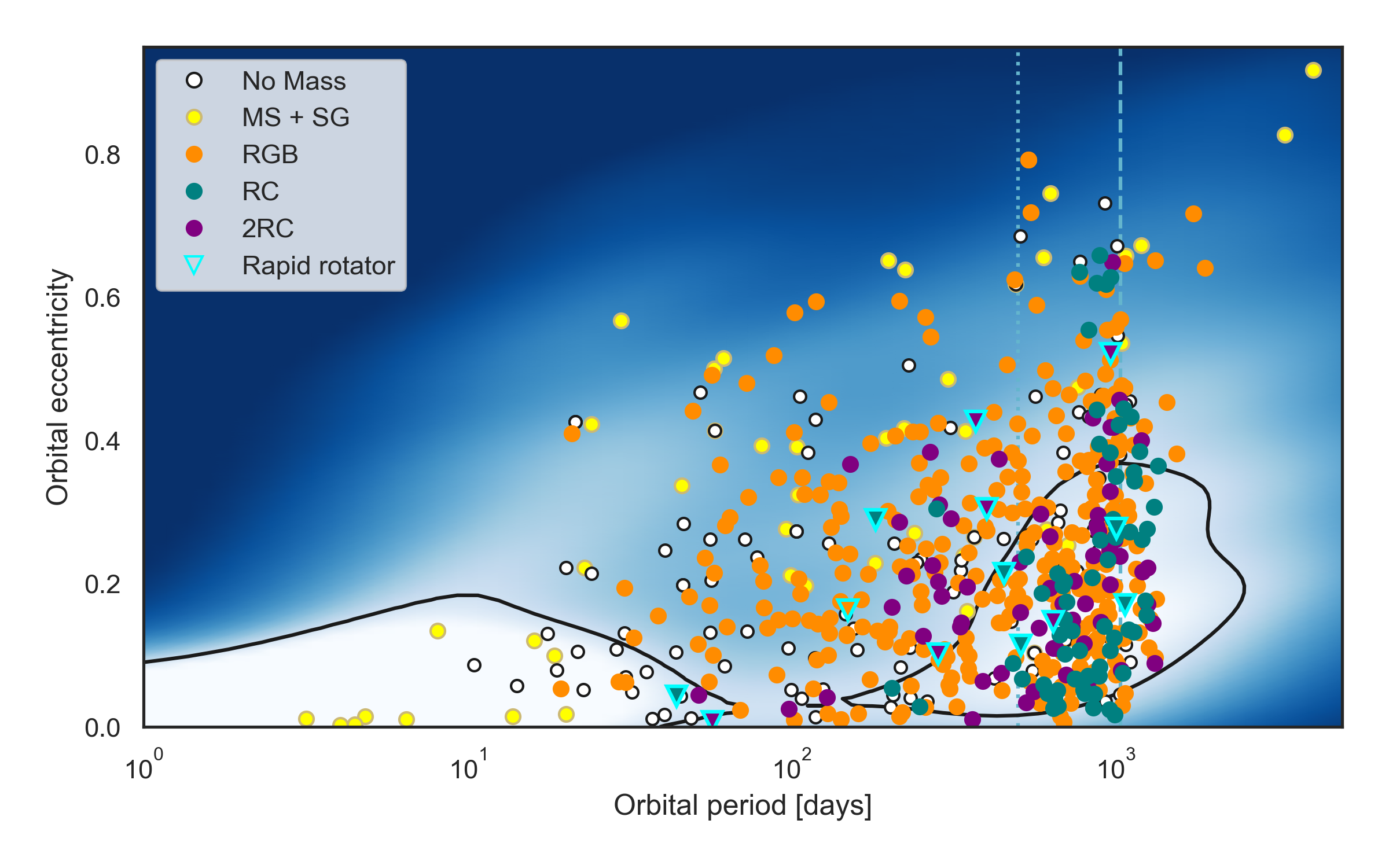}\vspace{-2mm}
    \caption{Orbital parameters of binary systems with seismically inferred evolutionary states. In addition to the color code for the evolutionary states of red giants (Fig.\,\ref{fig:seismicHRD}), oscillating MS+SG primaries are shown in yellow. Stars found to be rapid rotators are marked as triangles.  The background color map represents the KDE of the full SB9 catalog, with black lines outlining regions where the density exceeds seven times the median value. 
    Vertical dashed and dotted lines represent the 1034\,d timebase of \Gaia DR3 and $\sim$500\,d limit for RGB stars with degenerate cores to evolve as single stars. 
    Likely artifact solutions for giants (P$_\mathrm{orb,TBO}$\,$\leq$\,10\,days) are shown with a lower opacity. 
\label{fig:epPlaneEvol} }\vspace{-2mm}
\end{figure*}

For intermediate periods (8\,$\lesssim$\,P$_\mathrm{orb}$\,[d]\,$\lesssim$\,10$^4$), orbital eccentricities $e$ of MS binaries follow a Maxwellian “thermal” distribution \citep{Zahn1989, Raghavan2010, MoeStefano2017}, indicating that the probability density increases linearly with $e$, i.e., systems with higher eccentricities are more common than low or circularized eccentricities which is still present in the eccentricities of MS stars in Fig.\,\ref{fig:epPlaneEvol}. %The distribution changes for stars in more advanced evolutionary states \citep{Beck2024}. 

\begin{figure}[t]
    \centering
    \includegraphics[width=\columnwidth]{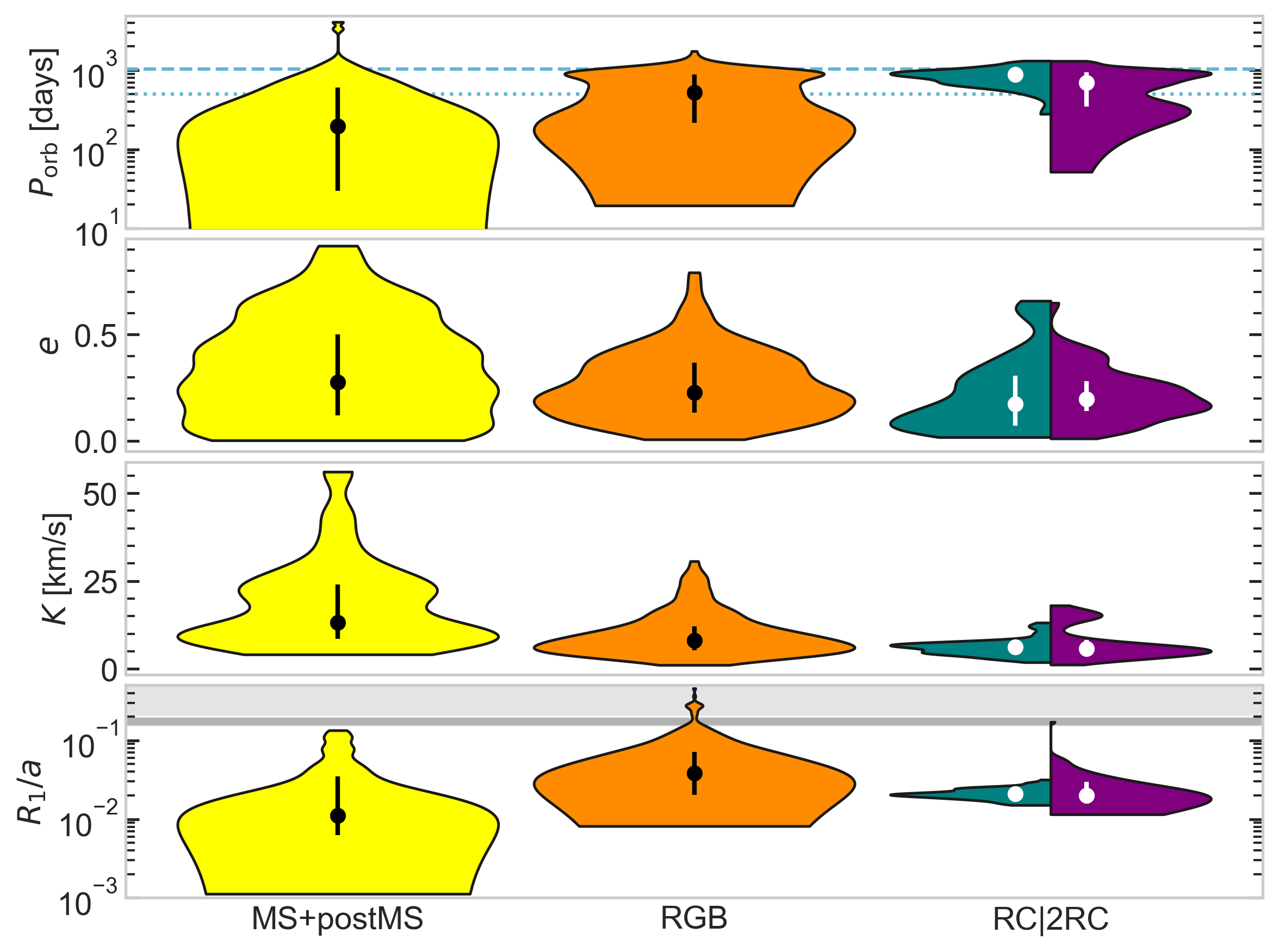}
    \caption{ KDEs of orbital parameters (e, P), the RV semi-amplitude $K$, and fractional radius $R_1/a$ for the binary population identified by the \Gaia TBO for solar-like oscillators. The distributions are grouped by evolutionary stage, with the rightmost violin in each panel split to show RC (teal) and 2RC (purple) stars. Dots and vertical bars indicate the median and interquartile range (25$^\mathrm{th}$–75$^\mathrm{th}$ percentiles) of each group. Horizontal dashed and dotted lines in the top panel mark the \Gaia\ DR3 time baseline (1034\,d) and the $\sim$500\,d upper limit for binary interaction during the RGB phase for stars with degenerate helium cores. The shaded gray region in the bottom panel marks the regime of strong tidal interaction ($R_1/a >$ 15\%).
    \label{fig:ePviolin} }\vspace{-2mm}
\end{figure}

The combined large sample of binary stars with orbital solutions from the TBO with seismically inferred evolutionary states and masses from the \apo and \apok allows us to populate the eccentricity-period (e-P) plane (Fig.\,\ref{fig:epPlaneEvol})  and study the binary evolution as a function of stellar evolution (Fig.\,\ref{fig:ePviolin}). 
We find a significant reduction in the ranges of orbital period and eccentricity from the MS into the He-core-burning phase. The short-periodic end of the period distribution shifts from a few days for \textit{MS} stars to about 20\,days for RGB. For stars in the He-core burning phase, the short periodic end is about 200 and 500\,days for 2RC and RC stars. This picture is consistent with the findings of \cite{DewberryWu2025}, who quantified comparable trends from a large-sample study of low-mass stellar binaries in the \GaiaDR catalog.

We further inspected the correlation between mass and orbital period as a function of the evolutionary state (Fig.\,\ref{fig:massPeriod}).  For MS and RGB stars, we find that the masses are mostly uncorrelated with period. However, for He-core burning stars, a clear inverse correlation is found between the primary mass and the orbital period of a system.
In particular, hardly any system with a primary mass M\,$\leq$\,1.8\,M$_\odot$ is found on orbital periods below 500\,days.
This supports the assumption that the difference in the cut-off periods results from the lower maximum radius at the RGB tip for stars with M\,$>$\,1.8\, M$_\odot$, allowing stars in smaller systems to complete the RGB phase without significant interactions.

Systems hosting helium-core burning stars show interesting dependencies of the eccentricity on their orbital period. 
Figure\,\ref{fig:epPlaneEvol} reveals that the sample of RC stars with P$_\mathrm{orb}$\,$\gtrsim$\,800\,days have a wide spread in eccentricity ($e$\,$\lesssim$0.7), similar to RGB stars, while for systems with 500\,$\lesssim$\,P$_\mathrm{orb}$\,$\lesssim$800\,days the eccentricity is  $e$\,$\lesssim$\,0.25. 
This step in the eccentricity distribution could suggest stronger tidal interactions up to this limiting period in systems hosting RC stars. The wider eccentricity range for the more massive 2RC stars ($e$\,$\lesssim$\,0.4) is explained by their smaller radii and weaker tidal interactions at He-core ignition. 
However, the steep increase and over-density of RC, 2RC, and RGB systems around 1,000\,days could also be partially an artifact. The comparison of \gaia TBO with SB9 by \citet[see Fig.\,2 and A.1]{Beck2024} showed that systems longer than the DR3 baseline of 1034\,days have periods around 1,000 days. The increased baseline in \gaia DR4, which is twice that of DR3, will help test whether this is an instrumental artifact.

\subsection{Post-common envelope systems}

\begin{figure}[t!]
    \centering
    \includegraphics[width=1.02\columnwidth]{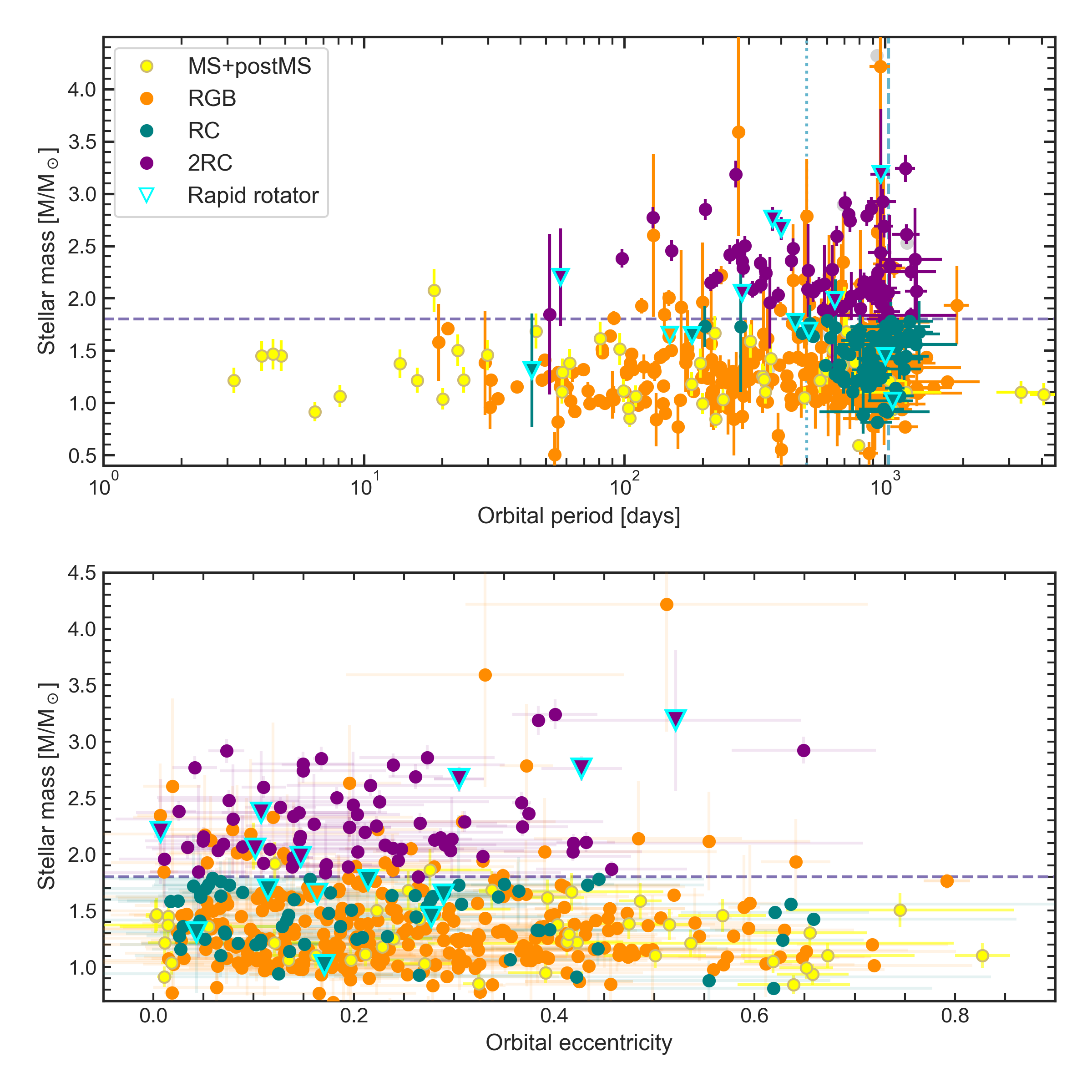}\vspace{-2mm}
    \caption{Orbital period and eccentricity as a function of the primary's mass. Colors, symbols, and the vertical dashed lines have identical meanings, as described in the caption of Fig.\,\ref{fig:epPlaneEvol}. The horizontal purple line marks the chosen separating mass between RC and 2RC. \vspace{-2mm}
    \label{fig:massPeriod} }
\end{figure}

We find numerous short periodic systems among the systems shown in the e-P plane in Fig.\,\ref{fig:epPlaneEvol}. Such systems are most likely artifacts of longer-periodic systems in the TBO. While red giants could exist in short-period systems, it is unlikely that such primaries would exhibit solar-like oscillations due to the effects of tidal interactions. We therefore excluded systems with P$_\mathrm{orb,TBO}$\,$\leq$\,10\ days from the analysis but show them in  Fig.\,\ref{fig:epPlaneEvol}.

Figures\,\ref{fig:epPlaneEvol} and \ref{fig:massPeriod} present several binary candidates from \Gaia TBO, which are hosting a He-core burning RC star at orbits significantly smaller than 500\,days. If such systems are real, these would be post-CE binary systems. When the secondary component does not spiral down deep enough to reach the bottom of the convective envelope, it is also possible that it survives in the convective envelope on an orbit smaller than the maximum radius of the giant at the tip of the RGB \citep{Ivanova2013}. It is striking that most RC primaries of these systems also have broadened absorption lines in APOGEE spectroscopy (triangle markers) rotationally. Such an enhanced surface rotation could indicate spin-up of the primary envelope during the CE phase, which was further enhanced during contraction as the star settled into the RC after helium ignition. 
Our ongoing spectroscopic follow-up observations will test if the orbital parameters from the TBO for those \hbox{particular systems are correct.}

While the mass distribution of rapidly rotating giants in H-shell burning RGB phase also follows the distribution of the entire sample (median: 1.21\,M$_\odot$, interquartile range: 1.03 to 1.53\,M$_\odot$), it strongly deviates in the He-core burning phase (median: 1.03\,M$_\odot$, interquartile range: 0.91 to 1.28\,M$_\odot$, Fig.\,\ref{fig:massHistograms}).
Such a difference supports the assumption that these stars have undergone a disruptive phase and lost larger fractions of their mass than would be accounted for from steady wind-driven mass loss, rendering ages from single-star models unreliable.

Only $\sim$7.5\% of all giants with enhanced surface rotation are found to be in binary systems. Although this is slightly above the overall binary rate for RC stars, it is notable, as \cite{Patton2024} reported, based on all available binary indicators from \Gaia DR3 and APOGEE RVs, a binary fraction of 73$\pm$2\% for the entire sample of giants with enhanced surface rotation. This is significantly higher than the expected binary fraction of low-mass MS stars, indicating that these are the products of mergers and disruptive events. 

\section{Evolution of the radial velocities \label{sec:RVevolution}}

The accumulated combined effect of stellar evolution and tidal interactions is propagating into the most commonly observed property of a binary system, the RV semi-amplitude, 
\begin{equation} 
K_{1,2} = \frac{M_{2,1}}{M_1 + M_2} \frac{(2\pi\,G)^{1/3}}{P_\mathrm{orb}^{1/3}} \frac{\sin i}{\sqrt{1 - e^2}},
\label{eq:RVsemiamplitude}
\end{equation} 
with $i$ being the inclination of the orbital plane. 
From classical mass loss and the tidally driven circularization of the orbit, a decrease in the RV amplitude is expected with continued age. 

From the evolution of $K$ with respect to \logg, \citet[Fig.\,5]{Badenes2018}  reported a significant attrition of the peak-to-peak amplitude from $\sim$200\,km/s on the MS to $\sim$30\,km/s at the RGB tip.
To test if we find a similar accumulated effect in the seismically characterized \apo and \apok samples, we used $K$ provided in the TBO \citep{Arenou2023}.  %It should be noted that this is a subset of the full TBO catalog. 

\begin{figure}[t!]\centering\vspace{-1mm}
    \includegraphics[width=0.999\columnwidth]{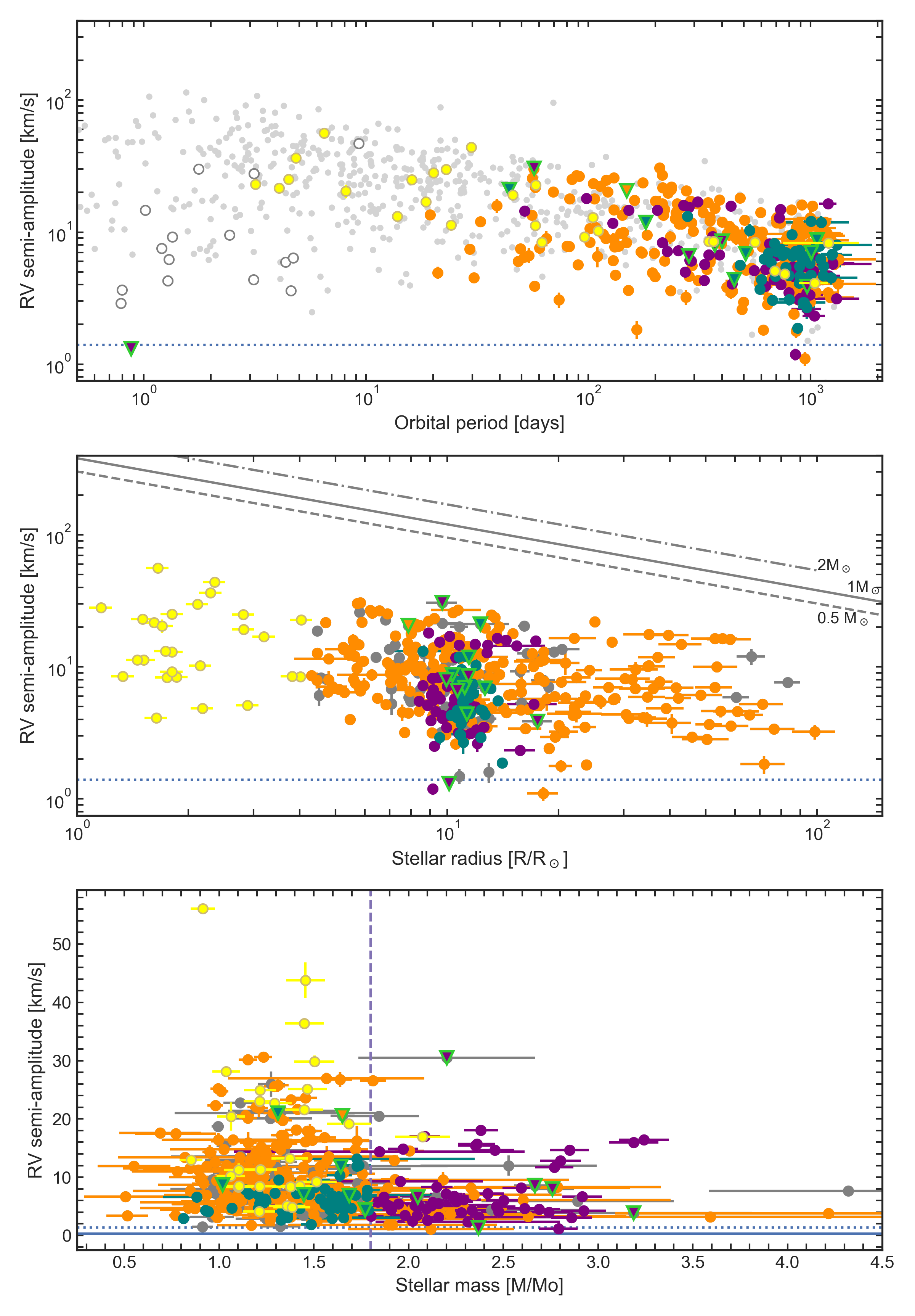}\vspace{-2mm}
    \caption{Evolution of the RV semiamplitude 
    of the binary motion as a function of the orbital period (top), stellar radius of the primary from \gaia parameters (middle), and the seismically inferred mass (bottom panel).
    The colors of the dots represent the various evolutionary states and are identical to Fig.\,\ref{fig:seismicHRD}.
    The dotted and solid horizontal lines in the top and bottom panels depict the detection threshold of the RVS instrument for faint (G\,$\leq$\,14\,mag) and bright stars (G\,$\leq$\,12\,mag), respectively.  
    The gray lines in the middle panel depict the RV for binary systems whose Roche lobe radius is filled. 
    %In the bottom panel, 
    The vertical purple line marks the chosen separation between RC and 2RC stars. 
\label{fig:semiAmplitudes} \vspace{-4mm}}
\end{figure}

\subsection{RV vs orbital period and stellar radius}

Figure\,\ref{fig:semiAmplitudes} (top panel, see also Fig.\,\ref{fig:ePviolin}) depicts an apparent reduction in the semi-RV amplitude from the 
MS is found. On the low-luminous RGB,  K\,$\lesssim$\,30\,km/s are found, but are reduced to K\,$\lesssim$\,20\,km/s on the luminous RGB. Most RC and 2RC systems are found with K\,$\lesssim$\,10\,km/s,  with a few systems \hbox{reaching  $K$\,$\lesssim$\,20\,km/s}. 

When comparing these numbers with \cite{Badenes2018}, we need to take into account that they are using the peak-to-peak variation, which is up to twice the semi-amplitude, K. Therefore, the values presented in Fig.\,\ref{fig:semiAmplitudes} are in good agreement with large-sample spectroscopic studies. The detection threshold value for a significant RV variation with the \gaia RVS instrument is 0.3 and 1.4\,km/s for bright and faint stars \citep{Katz2022}. Given the typical $K$ in red-giant binary systems and the brightness of targets in the \Kepler space photometry, this RV floor is sufficient.
Similarly, a reduction of $K$ is also seen as a function of the stellar radius (Fig.\,\ref{fig:semiAmplitudes}). Figure\,\ref{fig:ePviolin} shows that we find a cutoff of the oscillating sample of RGB stars at a fractional radius of $R_1/a$\,$\gtrsim$\,20\%. We further note that MS+SG and the core-He burning (RC+2RC) stars are well below this threshold (Fig.\,\ref{fig:ePviolin}).

The reduction of $K$ for RC and 2RC systems can be interpreted as a combined consequence of orbital circularization on the RGB, when the expanded envelope leads to large fractional radii and strong tides, and the effect of wind-driven mass loss, enhanced by the extreme stellar luminosities near the RGB tip.

To test if red-giant primaries are close to filling their Roche lobes, we estimated the critical period, P$_\mathrm{RL}$. We followed the formalism of \cite{Eggleton1983},
\begin{equation}
P_\mathrm{RL} = 2 \pi \cdot \mathcal{R}(q) \cdot \sqrt{\frac{\text{R}^3}{G M}},
\label{eq:PRL}
\end{equation}
where 
$\mathcal{R}(q)$ is a function of the mass ratio that gives the Roche lobe radius relative to the orbital separation, and $q$ is for the mass ratio, which we assume to be at unity. Following \cite{Eggleton1983}, $\mathcal{R}(q=1)$ can be approximated by 0.38 for the unity mass ratio. \cite{Miglio2014} pointed out that within a deviation of a few percent, $q$\,$\simeq$1 is expected for the majority of red-giant binary stars. Furthermore, \cite{Eggleton1983} noted that for 0.01\,$\leq$\,q\,$\leq$\,1, the function varies by less than 3\%. 
The resulting velocities for giants with 0.5, 1, and 2\,M$_\odot$, would be filling their Roche lobes are calculated from Eq.\,\ref{eq:RVsemiamplitude} and Eq.\,\ref{eq:PRL}, assuming an edge-on inclination ($i$\,=\,90) and circular orbits.
The gap between the detected systems and the Roche-lobe limit can be attributed to the suppression of modes due to enhanced activity driven by the increasing strength of the equilibrium tide. Therefore, those systems are far from filling their Roche lobe.

\subsection{RV versus stellar mass}

For He-core burning stars, no specific trend is found with mass, except for a few outliers ($K$\,$\lesssim$\,20\,km/s), in particular among the 2RC. An interesting trend is observed for the RGB stars, where the range for stars with M\,$\leq$\,1.8\,M$_\odot$ reaches up to 30\,km/s, while for systems with a more massive primary we find a similar range of RVs as for helium-core-burning stars. The RGB stars with the large $K$ are typically in shorter-periodic binaries (P$_\mathrm{orb}$\,$\lesssim$\,300\,days) which leads to higher $K$ values (Eq.\,\ref{eq:RVsemiamplitude}).

\section{Discussions and conclusions \label{sec:DiscussionConclusions}}
In this paper, we investigated the binary properties of red-giant stars by combining asteroseismic constraints from the APOKASC 3 and APO-K2 catalogs with binary solutions from \Gaia DR3’s NSS catalog. The asteroseismically inferred stellar masses and evolutionary states provide key diagnostics for characterizing binary systems hosting solar-like oscillators. This cross-matched dataset enabled a comprehensive analysis of how binary fraction, stellar mass distribution, and orbital parameters evolve across different evolutionary stages, offering a holistic view of stellar–binary coevolution.

Our results demonstrate a significant reduction in the binary fraction as stars ascend the RGB and transition into the core-helium-burning phase. To account for distinct evolutionary pathways, we divided the sample based on asteroseismic masses into low-mass stars (M\,$\leq$\,1.8\,M$_\odot$), which develop fully degenerate helium cores, and intermediate-mass stars (M\,$>$\,1.8\,M$_\odot$), which ignite helium under non-degenerate conditions. We find that approximately $\sim$31.5\% and $\sim$40.7\% of solar-like stars on the MS, with detected and non-detected solar-like oscillations, respectively, are members of binary systems. 
This is a noteworthy result, as magnetic activity, often enhanced in close binaries, has been suggested to reduce solar-like oscillations in dwarfs. Using $\nu_\mathrm{max}$ as a proxy for stellar luminosity, we further subdivided the RGB into low- and high-luminosity regimes. Our analysis reveals a significant reduction of $\sim$69\% in the binary fraction from the MS to the low-luminosity RGB, followed by a further $\sim$38\% decrease from the low- to high-luminosity RGB. Interestingly, no significant change in the binary fraction is found between the high-luminosity RGB and RC, suggesting that most binary attrition or interaction occurs earlier during the RGB ascent.
For the intermediate-mass stars (M\,$>$\,1.8\,M$_\odot$), we find a substantially higher binary fraction. This is consistent with the intrinsically higher binary occurrence rate (approx. 8 to 14\%), observed among early F- and A-type stars on the MS. However, due to the smaller sample size in this mass regime, we are currently unable to robustly test for evolutionary changes in binary fractions across the MS.

From the comparison of the mass of binaries hosting RGB and RC and 2RC, we find that the mass distribution for He-core burning stars is significantly shifted to more massive stars. This suggests that mass transfer, envelope stripping, or merger events have shaped a significant portion of the population and introduced a mass-dependent bias in binary attrition.

The distribution of orbital parameters reveals a clear evolutionary trend. While MS stars span a wide range of orbital periods, resembling the expected initial distribution, systems hosting RGB and core-helium-burning stars are predominantly found at longer periods and with reduced eccentricities. RC stars are typically found in systems with orbital periods longer than 500 days. In contrast, more massive 2RCs are found at shorter periods, down to $\sim$200 days. This reduction is interpreted as evidence of a binary interaction or disruption and is consistent with the fact that intermediate-mass stars ignite He at smaller radii on the RGB, allowing tighter systems to avoid interaction.

A few RC systems are detected with orbital periods shorter than 500 days, several of which exhibit signs of rapid surface rotation. These may represent candidates for post-CE systems or merger remnants resulting from previous mass-transfer episodes. Such systems exemplify the rich variety of products associated with binary interaction and mass transfer, such as hot subdwarf B stars \citep[sdB,][]{Vos2020}, He-rich white dwarfs \citep{Kilic2025}, binarity among central stars of planetary nebulae \citep{JonesBoffin2017}, symbiotic systems \citep[e.g.,][]{Merc2024}, where an evolved giant transfers material to a compact companion, and chemically peculiar barium stars \citep[e.g.,][]{Escorza2020}. From the reduction of the RV semi-amplitudes, we also show the accumulated effect of tidal interaction and mass loss. 

As orbital periods approach $\sim$800 to 1,000 days, corresponding to semimajor axes comparable to red-giant radii, binary coevolution becomes increasingly complex. 
In this regime, the otherwise simplifying assumption of coevolution of both stellar components and the binary orbit is challenged.
In particular, age estimates from single-star models may no longer be reliable for post-interaction products.

Our results provide clear observational evidence of mass- and evolution-dependent attrition of binary systems during the red-giant phase. These findings have important implications for binary population synthesis and the interpretation of chemical and rotational anomalies in evolved stars. In particular, the ages of helium-core burning stars may be misleading if prior mass loss or transfer is unaccounted for. Red-clump systems that have undergone substantial interaction are therefore unsuitable for calibrating stellar evolution models based on single-star physics. 
This study underscores the powerful synergy between asteroseismology and \Gaia in unraveling stellar and orbital coevolution, by providing valuable insights for more refined binary population synthesis and age-dating techniques.

The size of available samples of oscillating red giants in binaries is expected to grow substantially with upcoming data releases and forthcoming space missions. The extended time baseline and epoch spectroscopy provided by \Gaia DR4 will significantly improve the census and precision of astrometric and spectroscopic binary solutions, particularly for long-period systems. The Science Calibration and Validation Plato Input Catalog (scvPIC) of ESA’s PLATO mission \citep[PLAnetary Transits and Oscillations of stars;][]{Plato2024} will further increase the number of oscillating stars in eclipsing and astrometric binary systems identified by \Gaia. Additional contributions will come from NASA’s Roman Space Telescope \citep{huber2023roman} and the Chinese Earth\,2.0 mission \citep{Earch2pointO2022}, which will extend the ensemble of solar-like oscillators across diverse regions of the Milky Way. 
Together, these missions will provide the necessary data to probe binary evolution in the context of galactic archaeology. 
This underscores the need for a holistic picture of stellar and binary coevolution. This is essential for accurately determining stellar ages and unlocking the full potential of red giants as tracers of Galactic history.

\section*{Data Availability}
Tables \ref{tab:A1}, and \ref{tab:A2} are only available in electronic form at the CDS via anonymous ftp to \href{http://cdsarc.u-strasbg.fr/}{cdsarc.u-strasbg.fr} (130.79.128.5) or via \href{http://cdsweb.u-strasbg.fr/cgi-bin/qcat?J/A+A/}{http://cdsweb.u-strasbg.fr/cgi-bin/qcat?J/A+A/}.

\begin{acknowledgements}
The author thanks the referee for their valuable comments, which helped improve the article.
The author thanks the people behind the ESA \Gaia, NASA \Kepler, NASA TESS, and APOGEE projects. PGB thanks Marc Pinsonneault, Santi Cassisi, Thomas Masseron, Joel Zinn, Savita Mathur, Diego Godoy Rivera, and Jaroslav Merc,
for fruitful discussions on the catalogs and the paper manuscript.
PGB acknowledges support by the Spanish Ministry of Science and Innovation with the \textit{Ram{\'o}n\,y\,Cajal} fellowship number RYC-2021-033137-I and the number MRR4032204, as well as the proyecto plan nacional \textit{PLAtoSOnG} (grant no. PID2023-146453NB-100, PI: Beck). This project was supported by ``UNDARK'' of the Widening participation and spreading excellence programme (project number 101159929, PI: Camalich). This work was supported by the \hbox{International Space Science Institute (ISSI) in Bern (project\,24-629).}\\
This work has made use of data from the European Space Agency (ESA) mission
\href{https://www.cosmos.esa.int/gaia}{\Gaia}, processed by the \href{https://www.cosmos.esa.int/web/gaia/dpac/consortium}{\Gaia
Data Processing and Analysis Consortium (DPAC)}. Funding for the DPAC
has been provided by national institutions, in particular the institutions
participating in the \Gaia Multilateral Agreement. %\\
Data collected with the \textit{Kepler} mission, obtained from the MAST data archive at the Space Telescope Science Institute (STScI) was used. 
STScI is operated by the Association of Universities for Research in Astronomy, Inc., under NASA contract NAS 5–26555. 
\textit{Software:} \texttt{Python} \citep{10.5555/1593511}, 
\texttt{numpy} \citep{numpy,Harris_2020},  
\texttt{matplotlib} \citep{4160265},  
\texttt{scipy} \citep{2020SciPy-NMeth},
\texttt{pandas} \citep{reback2020pandas, mckinney-proc-scipy-2010},
\texttt{Astroquery} \citep{Grinsburg2019_astroquery}.
\end{acknowledgements}

\bibliographystyle{aa}
\bibliography{aa55157-25}

%\tableofcontents
%\end{document}
\begin{appendix}

\section{Table of orbital and seismic values \label{App:A}}

\begin{figure*}[t!]
    \vspace{-1mm}
    \centering
    \includegraphics[width=1.55\columnwidth]{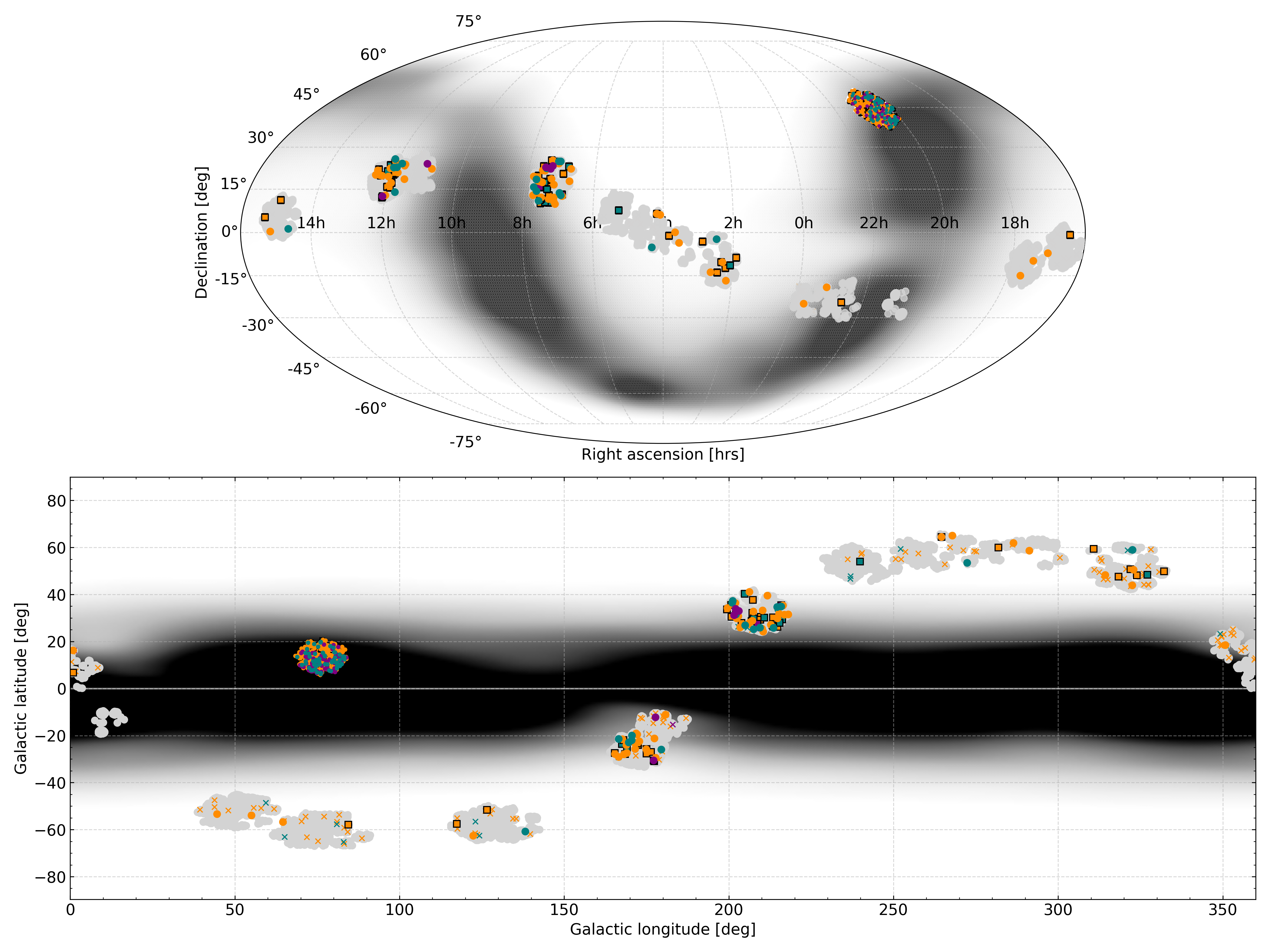}
\caption{Position of the observed samples on the equatorial plane of the sky (top panel)  as well as in galactic coordinates (bottom panel). 
Gray data points indicate the position of the individual targets in the catalog with detected oscillations. The K2 campaign (C) number is given. The shapes of the markers, representing various binary indicators, and their colors indicating seismically inferred evolutionary states are identical to those in Fig.\,\ref{fig:seismicHRD}.
The color-shaded background depicts the distribution of all stars observed by \gaia with $G$\,$\leq$\,10\,mag.
 \label{fig:galacticPlane} }
%\vspace{-3mm} 
\end{figure*}
\subsection{Notes on CDS Table: \GaiaDR binary systems in the APOKASC3 \label{tab:A1}}

The following list describes the parameters provided in Table~\ref{tab:A1}, which contains the orbital and asteroseismic properties of binary systems in the APOKASC-3 sample \citep{Pinsonneault2025} that have orbital solutions or other binary indicators from \GaiaDR \citep{Arenou2023}.

\begin{itemize}
    \item \verb|KIC_ID|: Kepler Input Catalog (KIC) identifier of the star. 

    \item \verb|nu_max|: Frequency of maximum oscillation power (\textmu Hz), derived from \textit{Kepler} photometry. 
    \item \verb|nu_max_err|: Uncertainty in the frequency of maximum oscillation power (\textmu Hz). 

    \item \verb|Delta_nu|: Large frequency separation between consecutive overtone modes (\textmu Hz). 
    \item \verb|Delta_nu_err|: Uncertainty in the large frequency separation (\textmu Hz). 

\item \verb|evo_state|: \verb|[MS_SG, RGB, RC, 2RC]| Evolutionary classification of the primary star, derived from asteroseismic diagnostics. 
The labels \texttt{MS}, \texttt{SG}, \texttt{RGB}, \texttt{RC}, and \texttt{2RC} correspond to main-sequence, subgiant, red-giant-branch, red-clump, and secondary-clump stars, respectively. 
The APOKASC-3 catalog originally provides evolutionary states only for giants (\texttt{RGB} and \texttt{RC}); here, the classification has been extended to include main-sequence (\texttt{MS}) and subgiant (\texttt{SG}) stars based on $\nu_\mathrm{max}$ and $\Delta\nu$ scaling relations and visual inspection of the seismic parameter space. For more details, we refer the reader to the description in Sects. \ref{sec:SampleDefinition} and  \ref{sec:BinaryFraction}.

    \item \verb|Teff|: Effective temperature (K) from APOGEE DR17. 
    \item \verb|Teff_err|: Uncertainty in the effective temperature (K). 

    \item \verb|[M/H]|: Spectroscopic metallicity (dex) from APOGEE DR17. 
    \item \verb|[M/H]_err|: Uncertainty in the metallicity (dex). 

    \item \verb|vsini|: Projected surface rotational velocity (km\,s$^{-1}$). 
    \item \verb|vsini_err|: Uncertainty in the projected rotational velocity (km\,s$^{-1}$). 

    \item \verb|Mass|: Stellar mass in solar units (M$_\odot$) inferred from asteroseismic scaling relations. 
    \item \verb|Mass_err|: Uncertainty in the stellar mass (M$_\odot$). 

    \item \verb|Radius|: Stellar radius in solar units (R$_\odot$) inferred from asteroseismic scaling relations. 
    \item \verb|Radius_err|: Uncertainty in the stellar radius (R$_\odot$). 
    \item \verb|sig_dr3|: Significance parameter of the Gaia TBO solution (Gaia DR3 NSS definition).
    \item \verb|Porb|: Orbital period (d) of the binary solution from the \GaiaDR3\ two-body orbit (TBO) catalog. 
    \item \verb|Porb_err|: Uncertainty in the orbital period (d). 

    \item \verb|e_orb|: Orbital eccentricity. 
    \item \verb|e_orb_err|: Uncertainty in the orbital eccentricity. 

    \item \verb|K1|: RV semi-amplitude (km\,s$^{-1}$) of the primary component derived from the Gaia orbital solution. 
    \item \verb|K1_err|: Uncertainty in the RV semi-amplitude (km\,s$^{-1}$). 

    \item \verb|RUWE|: Renormalized unit weight error from the \GaiaDR astrometric solution; indicator of astrometric excess noise. 

    \item \verb|NLAC|: \verb|[0,1]| Flag identifying if a source is listed (Flag=1) or not-listed (Flag=0) with nonlinear or acceleration solutions in the Gaia NSS catalog. 

    \item \verb|BUF|: \verb|[0,1]| Flag identifying if a source is listed (Flag=1) or not-listed (Flag=0) as ginary in the Binary Union Flag (BUF) \citep{GodoyRivera2025}. 

\end{itemize}

\newpage
\subsection{Notes on CDS Table: \GaiaDR binary systems in the APO-K2 \label{tab:A2}}

The following list describes the parameters provided in Table~\ref{tab:A1}, which contains the orbital and asteroseismic properties of binary systems in the APOKASC-3 sample \citep{Zinn2022} that have orbital solutions or other binary indicators from \GaiaDR \citep{Arenou2023}.

\begin{itemize}
    \item \verb|EPIC_ID|: Ecliptic Plane Input Catalog (EPIC) identifier of the \textit{Kepler}/\textit{K2} target star.

    \item \verb|nu_max|: Frequency of maximum oscillation power (\textmu Hz), derived from \textit{K2} photometry \citep{Zinn2022}. 
    \item \verb|nu_max_err|: Uncertainty in the frequency of maximum oscillation power (\textmu Hz). 

    \item \verb|Delta_nu|: Large frequency separation between consecutive overtone modes (\textmu Hz), derived from asteroseismic analysis of \textit{K2} light curves. 
    \item \verb|Delta_nu_err|: Uncertainty in the large frequency separation (\textmu Hz). 

    \item \verb|evo_state|: \verb|[MS_SG, RGB, RC, 2RC]| Evolutionary classification of the primary star, derived from asteroseismic diagnostics. 
    The labels \texttt{MS}, \texttt{SG}, \texttt{RGB}, \texttt{RC}, and \texttt{2RC} correspond to main-sequence, subgiant, red-giant-branch, red-clump, and secondary-clump stars, respectively. 
    The classification for the \textit{K2} targets follows the same method used for the APOKASC-3 sample, based on $\nu_\mathrm{max}$ and $\Delta\nu$ scaling relations and inspection of the seismic parameter space (see Sects.~\ref{sec:SampleDefinition} and~\ref{sec:BinaryFraction}).

    \item \verb|Teff|: Effective temperature (K) from APOGEE~DR16 or GALAH~DR2 spectroscopy. 
    \item \verb|Teff_err|: Uncertainty in the effective temperature (K). 

    \item \verb|[M/H]|: Spectroscopic metallicity (dex) from APOGEE~DR16 or GALAH~DR2. 
    \item \verb|[M/H]_err|: Uncertainty in the metallicity (dex). 

    \item \verb|Mass|: Stellar mass (M$_\odot$) inferred from asteroseismic scaling relations using $\nu_\mathrm{max}$, $\Delta\nu$, and $T_\mathrm{eff}$. 
    \item \verb|Mass_err|: Uncertainty in the stellar mass (M$_\odot$). 

    \item \verb|Radius|: Stellar radius (R$_\odot$) inferred from asteroseismic scaling relations. 
    \item \verb|Radius_err|: Uncertainty in the stellar radius (R$_\odot$). 

    \item \verb|sig_dr3|: Significance parameter of the Gaia TBO solution (Gaia DR3 NSS definition).
    \item \verb|Porb|: Orbital period (d) of the binary solution from the \GaiaDR3\ two-body orbit (TBO) catalog. 
    \item \verb|Porb_err|: Uncertainty in the orbital period (d). 

    \item \verb|e_orb|: Orbital eccentricity. 
    \item \verb|e_orb_err|: Uncertainty in the orbital eccentricity. 

    \item \verb|K1|: RV semi-amplitude (km\,s$^{-1}$) of the primary component derived from the Gaia orbital solution. 
    \item \verb|K1_err|: Uncertainty in the radial-velocity semi-amplitude (km\,s$^{-1}$). 

    \item \verb|RUWE|: Renormalized unit weight error from the \GaiaDR\ astrometric solution; indicator of astrometric excess noise. 

    \item \verb|NLAC|: \verb|[0,1]| Flag indicating whether a source is listed (\texttt{1}) or not listed (\texttt{0}) with nonlinear or acceleration solutions in the Gaia NSS catalog. 
\end{itemize}

\setcounter{table}{2}
\begin{table*}[th!]
    \centering
\tabcolsep=10pt
\caption{\GaiaDR solution of binary systems with the primary's projected surface rotation above 5 km/s. }
%\vspace{-2mm}
    \begin{tabular}{rrrc|rrrrrrr}
\hline\hline

\multicolumn{1}{c}{KIC} &
\multicolumn{1}{c}{Mass} & 
\multicolumn{1}{c}{$\sigma$(Mass)} & 
\multicolumn{1}{c}{Evol.} & 
\multicolumn{1}{c}{$P_{\rm orb}$} &  
\multicolumn{1}{c}{$\sigma(P_{\rm orb})$} &  
\multicolumn{1}{c}{$e$} &  
\multicolumn{1}{c}{$\sigma(e)$} &        
\multicolumn{1}{c}{Signif.} &       
\multicolumn{1}{c}{$v\sin i$} & 
\multicolumn{1}{c}{$G$}\\

\multicolumn{1}{c}{} &
\multicolumn{1}{c}{[M/M$_\odot$]} &
\multicolumn{1}{c}{[M/M$_\odot$]} &
\multicolumn{1}{c}{State} &
\multicolumn{1}{c}{[days]} &
\multicolumn{1}{c}{[days]} &
\multicolumn{1}{c}{} &
\multicolumn{1}{c}{} &
\multicolumn{1}{c}{DR3} &
\multicolumn{1}{c}{[km/s]} &
\multicolumn{1}{c}{[mag]}
 \\[1mm] \hline	
 
 2443035 & 1.64 &   0.07 &  RC &             181.61 &                      0.82 & 0.29 &         0.05 &         13.4 &    5.2 &     12.2 \\
 5290071 & 1.45 &   0.06 &  RC &            1005.83 &                     66.92 & 0.28 &         0.12 &         13.6 &    5.9 &     12.4 \\
 7500236 & 1.31 &   0.54 &  RC &              44.13 &                      0.03 & 0.04 &         0.04 &         30.7 &   10.0 &     12.5 \\
 9086060 & 1.77 &   0.06 &  RC &             453.20 &                     11.43 & 0.21 &         0.09 &         11.8 &    6.6 &     11.9 \\
10355764 & 1.02 &   0.04 &  RC &            1069.79 &                    209.57 & 0.17 &         0.11 &         10.0 &    5.6 &     12.3 \\
11134475 & 1.69 &   0.07 &  RC &             511.03 &                     17.61 & 0.11 &         0.10 &         10.0 &    5.5 &     12.8
\\[1mm] \hline	

3458643 & 2.67 &   0.11 & 2RC &             399.53 &                      1.69 & 0.31 &         0.02 &         37.4 &    5.6 &     10.2 \\
 3458643 & 2.67 &   0.11 & 2RC &             399.53 &                      1.69 & 0.31 &         0.02 &         37.4 &    5.6 &     10.2 \\
 6516800 & 2.05 &   0.08 & 2RC &             283.11 &                      2.67 & 0.10 &         0.05 &         18.7 &    7.4 &     10.9 \\
 6866251 & 2.76 &   0.11 & 2RC &             370.64 &                      3.20 & 0.43 &         0.04 &         17.7 &    5.1 &     11.5 \\
 7103951 & 1.98 &   0.08 & 2RC &             642.64 &                      2.50 & 0.15 &         0.01 &         71.6 &    6.3 &      9.3 \\
 %7670875 & 2.37 &   0.55 & 2RC &               0.88 &                      0.00 & 0.11 &         0.13 &          7.3 &    5.8 &     10.1 \\
11297585 & 3.19 &   0.62 & 2RC &             963.27 &                     83.37 & 0.52 &         0.13 &         14.7 &    6.7 &     10.1 \\
11600413 & 2.20 &   0.47 & 2RC &              56.94 &                      0.01 & 0.01 &         0.01 &        189.4 &    7.0 &     10.2

\\[1mm] \hline	

10796857 & 1.65 &   0.06 & RGB &              149.1 &                      0.54 & 0.16 &         0.04 &         19.6 &    6.2 &     12.4 
\\[1mm] \hline	
\end{tabular}
    \tablefoot{See appendix \ref{App:A3} for for a detail description of the table's content.} \vspace{-3mm}
    \label{tab:rapidRotBinaries} 
\end{table*}

\newpage
\subsection{Systems hosting rapidly rotating primaries \label{App:A3}}

The following list describes the columns presented in Table\,\ref{tab:rapidRotBinaries}.

\begin{itemize}
\item {KIC}: Kepler Input Catalog identifier.

\item {Mass} \,[M/M$_\odot$]: Stellar mass inferred from asteroseismic scaling relations.
\item {$\sigma$(Mass)} \,[M/M$_\odot$]: Uncertainty of the stellar mass.

\item {Evol.\,State}: Asteroseismic evolutionary state of the primary (\texttt{RGB}, \texttt{RC}, \texttt{2RC}).

\item \textbf{$P_{\rm orb}$} \,[days]: Orbital period of the Gaia DR3 two–body orbit (TBO) solution.
\item \textbf{$\sigma(P_{\rm orb})$} \,[days]: Uncertainty of the orbital period.

\item {$e$}: Orbital eccentricity of the Gaia TBO solution.
\item {$\sigma(e)$}: Uncertainty of the orbital eccentricity.

\item {Signif.}: Significance parameter of the Gaia TBO solution (Gaia DR3 NSS definition).

\item {$v\sin i$} \,[km\,s$^{-1}$]: Projected surface rotational velocity (from APOGEE DR17).

\item {$G$} \,[mag]: Gaia broad–band G magnitude.
\end{itemize}

\end{appendix}
\end{document}